\documentclass[12pt,letterpaper,nofootinbib,preprint,superscriptaddress,titlepage]{revtex4}
\usepackage[top=1in, bottom=1in, left=1in, right=1in, footskip=0.5in, headheight=0.5in, footnotesep=0.2in, marginparwidth=0in, marginparsep=0in]{geometry}

\usepackage{amsmath,amsthm,amssymb}
\usepackage{esvect,graphicx,latexsym,eso-pic,color}

\definecolor{darkblue}{rgb}{0.0, 0.0, 0.55}
\definecolor{forestgreen}{rgb}{0.13, 0.55, 0.13}
\usepackage[colorlinks=true,linktocpage=true,linkcolor=darkblue,citecolor=forestgreen,urlcolor=forestgreen]{hyperref}

\usepackage[hang,flushmargin]{footmisc}
\usepackage{placeins}
\usepackage{bbold}
\usepackage{multirow}
\usepackage{slashed}
\usepackage{cancel}

\DeclareMathOperator{\gev}{GeV}

\DeclareMathOperator{\fb}{fb}

\DeclareMathOperator{\Br}{\mathrm{Br}}

\newcommand{\beq}{\begin{equation}} \newcommand{\eeq}{\end{equation}}
\newcommand{\bea}{\begin{eqnarray}} \newcommand{\eea}{\end{eqnarray}}

\DeclareMathOperator{\br}{Br}

\begin{document}
\preprint{YITP-SB-21-1}

\title{Multi-Higgs Production Probes Higgs Flavor}

\author{Daniel Egana-Ugrinovic}
\affiliation{Perimeter Institute for Theoretical Physics, Waterloo, ON, N2L 2Y5}
\author{Samuel Homiller}
\affiliation{Department of Physics, Harvard University, Cambridge, MA, 02138}
\author{Patrick Meade}
\affiliation{C. N. Yang Institute for Theoretical Physics, Stony Brook University, Stony Brook, NY 11794}

\begin{abstract}

We demonstrate that multiple-Higgs production at the LHC is the most sensitive probe of first and second-generation quark flavor in the Higgs sector.
In models where new scalars couple to light quarks, 
gigantic di-Higgs and even sizable tri-Higgs production rates can be obtained, 
which can be used to either discover or severely constrain such theories.
As an example, 
we show that the most stringent bounds on enhanced interactions of the $125\,\textrm{GeV}$ Higgs to the down quark in extended Higgs sectors 
are obtained by looking for the extra Higgs bosons that provide for such enhancements using the di-Higgs and $Zh$ topologies. 
In this context, we set new limits on the 125 GeV Higgs coupling to the down quark as strong as $\lambda_{hd\bar{d}} \lesssim 30 \lambda_{hd\bar{d}}^{\textrm{SM}}$ --- a dramatic improvement over previously available bounds.
Regarding second-generation quark flavor, we obtain new limits in the coupling to strange as strong as $\lambda_{hs\bar{s}} \lesssim 10 \lambda_{hs\bar{s}}^{\textrm{SM}}$.
In addition, we show that the currently unexplored triple-Higgs production topology could be a potential discovery channel of a wide variety of extended Higgs sectors at the LHC, including not only models where extra Higgses couple to light quarks, but also popular theories where they have preferential couplings to the the top.
\end{abstract}
\maketitle

\tableofcontents

\newpage
\section{Introduction}
The Higgs sector of the Standard Model (SM) is central to a preponderance of unanswered questions in particle physics.   Despite our growing knowledge of Higgs properties from the LHC, many remain unknown even at an $\mathcal{O}(1)$ level of uncertainty. If the SM is the correct theory of the Electroweak (EW) scale, then some properties of the Higgs will remain out of reach of the HL-LHC or even proposed Higgs factories. 
A paramount example of this is the flavor sector, where there are not even ideas of how to measure the light quark Yukawas in the SM at the $\mathcal{O}(1)$ level.
Other aspects, such as the shape of the Higgs potential itself, 
require new Energy Frontier machines beyond the LHC to achieve percent-level precision.  Nevertheless, flavor and the Higgs potential are two of the most interesting aspects of the Higgs, and are both highly motivated windows into physics beyond the SM (BSM).

Modifications to the Higgs potential and flavor sector are normally thought of as distinct topics, pointing to dramatically different energy scales.
Flavor physics, thought of from the agnostic, bottom-up EFT perspective, points to scales well above any proposed collider based on current experimental data. Changes to the Higgs potential, on the other hand, are far less constrained, and believed to be tied close to the EW scale based on motivations such as naturalness or the EW phase transition (EWPT)~\cite{Dimopoulos:1981zb,Cline:2006ts,Trodden:1998ym,Curtin:2014jma,Mazumdar:2018dfl}.
Observables related to the Higgs potential and flavor are often thought of as quite disparate as well.
In the case of the Higgs potential, multi-Higgs production, specifically di-Higgs production via the triple-Higgs coupling, is viewed as the best probe given its cross section and correlation to the potential~\cite{DiMicco:2019ngk}.
Meanwhile, non-trivial flavor associated with the Higgs is often relegated to third generation decays or associated production of a single Higgs, both from the perspective of weaker flavor bounds and flavor tagging.
It turns out, though, that these two topics can be tightly intertwined.
In particular, we will show that multi-Higgs production currently sets the {\em strongest} bounds on modifications to the Higgs couplings to light quarks when obtained in theories of extended Higgs sectors.
Conversely, we will demonstrate that flavorful theories at low scales can significantly enhance multi-Higgs production rates far beyond most other BSM theories, to the point that the HL-LHC has significant discovery potential.

That flavor and multi-Higgs production can be closely linked is due to the fact that multi-Higgs production is not governed solely by the Higgs potential parameters. This is well known in the context of di-Higgs production in the SM, where small shifts in the top Yukawa can offset shifts in the triple Higgs coupling~\cite{Goertz:2014qta}.
This can be understood generically from the fact that the production cross section can be controlled by different couplings than those governing the multi-Higgs final state.
In particular, if physics at the TeV scale is flavored (i.e., has couplings that aren't universal to the fermions), there can be new production mechanisms for multi-Higgs final states such as quark fusion that dramatically change the experimental prospects.

While flavorful new physics and multi-Higgs production can be linked, this relation has not been discussed in past literature because of the aforementioned experimental bounds on bottom up EFTs with generic flavor structures~\cite{Bona:2007vi}, and also due to theoretical bias on the flavorful couplings that new physics near the EW scale may have~\cite{DAmbrosio:2002vsn}.
To overcome these issues,
in this paper we build on our previous work, Spontaneous Flavor Violation (SFV)~\cite{Egana-Ugrinovic:2018znw,Egana-Ugrinovic:2019dqu}, which is a UV complete framework that allows new physics at the EW scale to have flavorful couplings while satisfying all flavor bounds. 
This framework provides strong motivation to study a set of simple theories:
extended Higgs sectors where extra Higgs bosons can have large couplings to light quarks (first or second-generation)~\cite{Egana-Ugrinovic:2019dqu}.
In these models, the connection between multi-Higgs production and flavor becomes clear, as extra Higgses are resonantly produced at hadron colliders via quark fusion, 
and since they do not carry conserved quantum numbers they naturally decay into two or three Higgs bosons. 
The whole process happens at tree-level so gigantic di-Higgs rates are obtained --- 
orders of magnitude larger than the SM predictions --- 
which are well within the reach of LHC searches. 
Large resonant tri-Higgs production rates can also be realized,
which could be used to discover extended Higgs sectors at the LHC if dedicated searches were performed. 
In addition, the $125\,\textrm{GeV}$ Higgs couplings to the light quarks are modified via mixing with the extra Higgses, so multi-Higgs bounds on the new states can be straightforwardly interpreted as strong (though model-dependent) bounds  on the Higgs couplings. 
In previous literature, modifications to the Higgs light-quark Yukawas have been studied in the context of EFTs~\cite{Bodwin:2013gca,Kagan:2014ila,Perez:2015lra,Zhou:2015wra,Brivio:2015fxa,Bishara:2016jga,Soreq:2016rae,Yu:2016rvv,Aaboud:2017xnb,Aaboud:2018fhh,Coyle:2019hvs,Sirunyan:2020mds,Falkowski:2020znk},
but in that case only very large couplings can be tested. 
Since large modifications can be obtained only if the new physics is near the electroweak scale, 
it is then natural to study the new physics states directly, as done here,
especially since there are very few models in which these modifications can be realized.

While our main focus is on theories where extra Higgses have sizable couplings to light quarks, 
we also find that tri-Higgs production can be realized at the LHC even in models where such couplings do not exist, and only couplings to the third-generation quarks are large, 
such as 
the type I-IV 2HDMs~\cite{Glashow:1976nt} or the singlet-extended SM~\cite{Veltman:1989vw,McDonald:1993ex,OConnell:2006rsp}.
In this case, tri-Higgs production happens at loop level
via the associated production of an extra Higgs with a $125\,\textrm{GeV}$ Higgs, with the extra Higgs subsequently decaying to two more $125\,\textrm{GeV}$ Higgses.
In summary, our work indicates several new and well-defined targets for both current and upcoming multi-Higgs searches, and provides strong motivation to perform novel dedicated tri-Higgs searches.

We organize this paper as follows.
To demonstrate the generic lessons that one can take away from the combination of BSM flavor and the Higgs, we begin with a general introduction to multi-Higgs production in Section~\ref{sec:simplified}.  We explore a generic range of BSM predictions for both di-Higgs and tri-Higgs production at a hadron collider, and explain why flavorful production is one of the most interesting channels for the LHC. We also show that even though one can describe flavorful production through a higher dimension operator akin to the SM EFT framework~\cite{Alasfar:2019pmn}, this does not capture the interesting physics possibilities for the LHC.

In Section~\ref{sec:2HDM} we review the UV complete theory in which extra scalars have large couplings to light quarks, introduced in ref.~\cite{Egana-Ugrinovic:2019dqu}.
We show the range of possibilities for di-Higgs and tri-Higgs production in this concrete BSM flavored model in Section~\ref{sec:dihiggs}, followed by the bounds on enhancements of the light Yukawa couplings for the SM Higgs in Section~\ref{sec:enhancements}.  
Finally, we demonstrate projections for the HL-LHC in Section~\ref{sec:projections} and point out directions for further detailed experimental searches, including tri-Higgs production as a probe of extended Higgs sectors both with and without large couplings to light quarks.

\section{Multi-higgs production in extended Higgs sectors: general features}
\label{sec:simplified}

\begin{figure}[t!]
\vspace{0.5cm}
\quad
\includegraphics[scale=0.4]{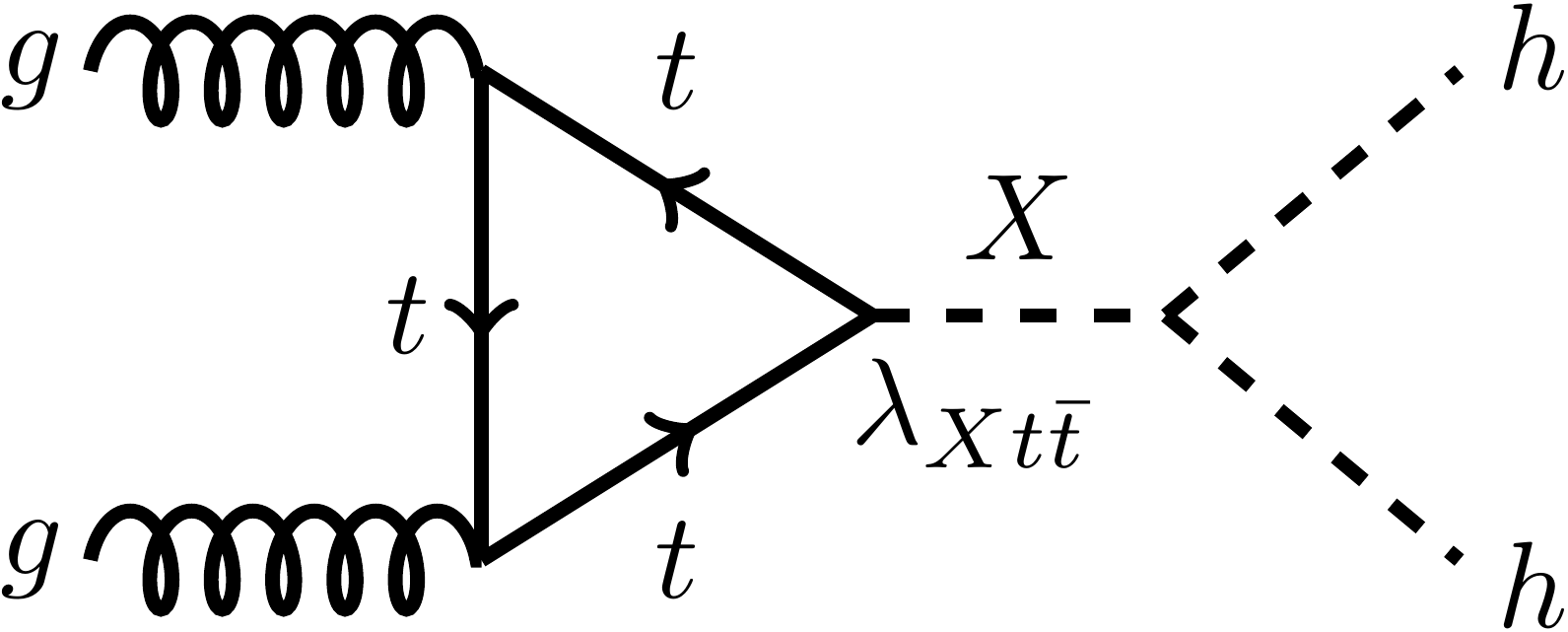}
\hspace{1cm}
\includegraphics[scale=0.4]{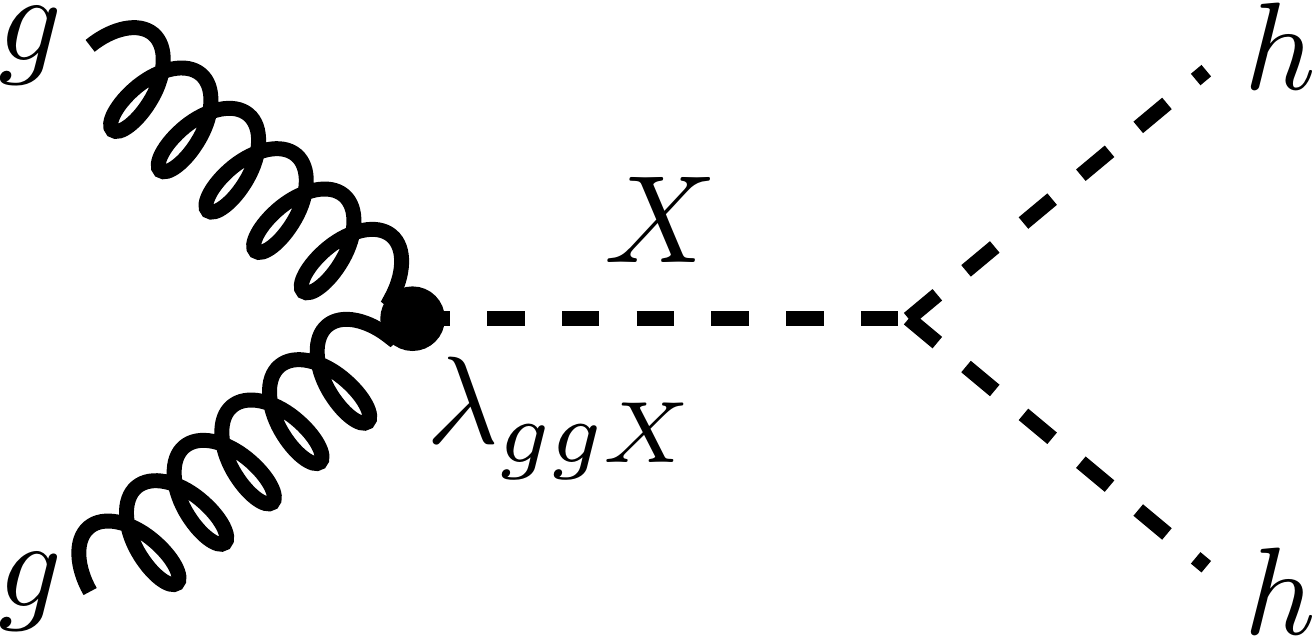}
\quad \\
\vskip 0.5cm
\includegraphics[scale=0.4]{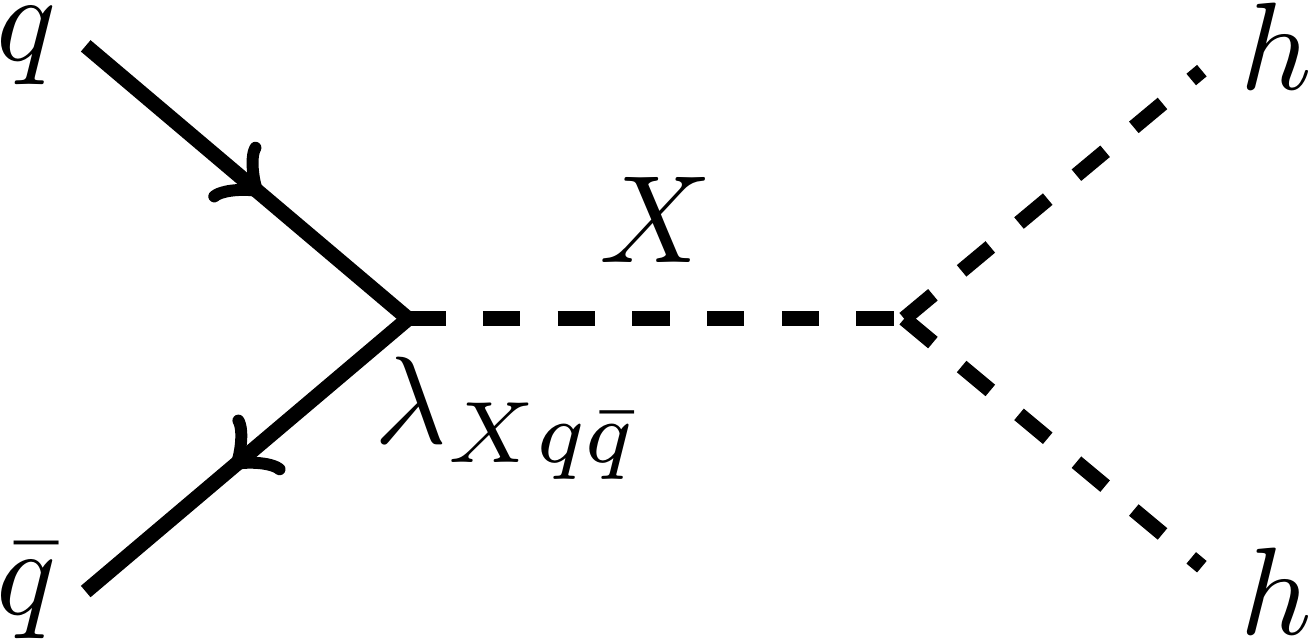}
\caption{Diagrams leading to resonant production of Higgs pairs at the LHC, via a scalar field $X$ coupling to gluons (upper left and right diagrams), or light quarks (bottom diagram).
Resonant tri-Higgs production is mediated by similar diagrams, with the addition of one Higgs boson to the rightmost vertices.}
\label{fig:diagrams}
\end{figure}

In this section we provide an overview of the discovery potential of multiple-Higgs production at the LHC in models of extended Higgs sectors.
Large di-Higgs or tri-Higgs rates can be naturally obtained when extra scalar states are resonantly produced, and subsequently decay to $125\,\textrm{GeV}$ Higgses.   To illustrate the panoply of possibilities we take the case of a neutral scalar $X$ (which may stand in for SM singlets, scalars within doublets, etc., and may be part of a larger sector) that is resonantly produced and then decays into a multi-Higgs final state.
In this case, the multi-Higgs cross section is simply given by the scalar's production cross section times the corresponding branching fraction into $125\,\textrm{GeV}$ Higgses,
$\sigma(hh)=\sigma_X \times \br(X\rightarrow h h)$ or $\sigma(hhh)=\sigma_X \times \br(X\rightarrow h hh)$.
At hadron colliders, $\sigma_X$  can be parametrized in terms of the interactions between the scalar and quarks or gluons,
which mediate di-Higgs production via the diagrams in Fig.~\ref{fig:diagrams},
or tri-Higgs production via similar diagrams.

The most familiar way for scalars to be produced at hadron colliders with large cross sections is through a coupling to top quarks, as in the case of the SM Higgs, which induces a coupling to gluons at one-loop (as in Fig.~\ref{fig:diagrams}), \footnote{The corresponding diagrams with lighter quarks in the loop are chirally suppressed.} 
\begin{equation}
\mathcal{L}_t = -\lambda_{Xt\bar{t}} \,\, X t \bar{t} + \textrm{h.c.}.
\label{eq:topcoupling}
\end{equation}
The top-quark Yukawa operator  arises in popular theories such as the singlet-scalar extension of the SM (``minimal Higgs portal'') \cite{Veltman:1989vw,McDonald:1993ex,OConnell:2006rsp}, the types I-IV 2HDMs \cite{Glashow:1976nt} and the Higgs sector of the MSSM \cite{Dimopoulos:1981zb}.
Going beyond the renormalizable level, scalars may also couple to gluons via the operator
\begin{equation}
\mathcal{L}_g = \lambda_{Xgg} \,\,  \frac{\alpha_s}{12\pi v} \, X G^{\mu\nu} G_{\mu\nu} ,
\label{eq:gluonoperator}
\end{equation}
which arises at one-loop in theories where gauge-singlet scalars couple to heavy vector-like quarks \cite{Nakamura:2017irk}.

A less investigated possibility is that a scalar may couple to light quarks ($q=u,d,s$) via tree-level Yukawa interactions that are larger than SM Yukawas,
\begin{equation}
\mathcal{L}_q = -\lambda_{Xq \bar{q}} \,\, X q \bar{q} + \textrm{h.c.}  , \qquad q=u,d,s . \\
\label{eq:quarkcoupling}
\end{equation}
The above operator arises at the renormalizable level in theories where the scalar is part of a second-Higgs doublet \cite{Egana-Ugrinovic:2019dqu}, or at the non-renormalizable level in singlet-scalar models with dimension-five interactions \cite{Batell:2017kty}.
The reason why this case is typically explored less, is because without a symmetry reason, such as SFV,  flavor constraints require very small $\lambda_{Xq \bar{q}}$.  However, as we will show in this agnostic description, the coupling does not have to be too large to have dramatic experimental consequences.

Once produced, decays of the scalar $X$ into SM Higgses can be obtained from a super-renormalizable operator with coupling strength $\xi$
\begin{equation}
\xi X h h ,
\label{eq:xicoupling}
\end{equation}
which arises in minimal Higgs portal models. Extra scalars can also decay to two and three Higgses via a quartic coupling $\lambda$
\begin{equation}
3 \lambda X v h h, 
\qquad
\lambda X h h h ,
\label{eq:quarticdecay}
\end{equation}
that exists in two-Higgs doublet models. 
Note that decays into three Higgses are only possible at the renormalizable level via the quartic coupling Eq.~\eqref{eq:quarticdecay}, which also allows for decays into two Higgses,
so a relation between these two decay channels is established.
In particular, given the higher multiplicity of the $3h$ final state, $\br(X\rightarrow h h h)$ is  generically expected to be smaller than $\br(X\rightarrow h h)$.

The branching fractions into $125\,\mathrm{GeV}$ Higgses are further specified by a variety of model-specific considerations, 
such as the availability of other decay modes and
vacuum stability (see e.g. \cite{Chen:2014ask, DiLuzio:2017tfn}).
However, in order to evaluate if resonant multi-Higgs production is a relevant probe of extended Higgs sectors 
it is illustrative to consider the maximal possible rates, which are obtained by working in the limit where the branching fraction into two Higgses is maximal, $\br(X\rightarrow h h) \sim 1$. 
In that case we can expect at most $\br(X\rightarrow h h h)\sim10^{-2}$ from phase space considerations.
Under these assumptions, the resonant di-Higgs production cross section can be approximated by the scalar's production cross section, $\sigma(hh) \sim \sigma_X$, while the resonant tri-Higgs cross section is expected to be at most $\sigma(hhh) \sim 10^{-2} \sigma_X$. 

\begin{figure}[htbp!]
\includegraphics[width=\linewidth]{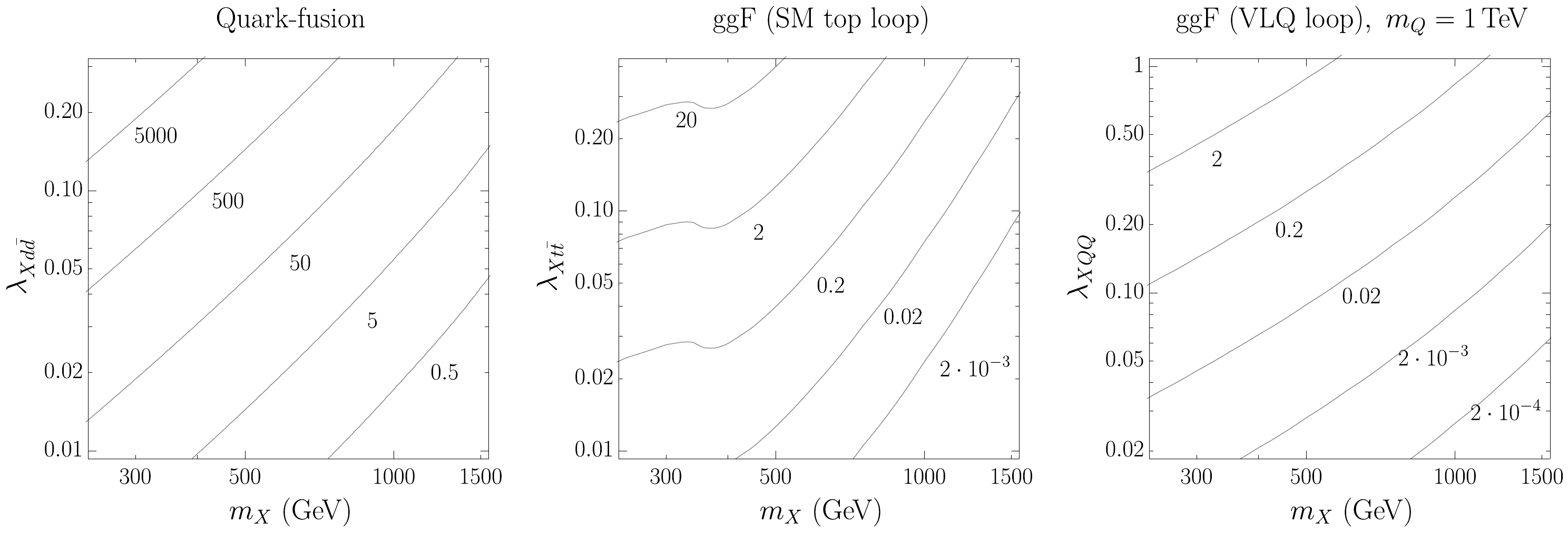}
\caption{Contours of the resonant scalar production cross section $\sigma_X$ at the 13 TeV LHC, as a function of the resonance mass $m_X$ and the effective couplings to first-generation down quarks (left), top quark (middle) and heavy vector-like quarks of mass $m_Q=1$\,TeV that generate a dimension-5 coupling to gluons (right). 
For the rightmost panel, the Yukawa coupling to vector-like quarks $\lambda_{XQQ}$ is related to $\lambda_{Xgg}$ in Eq. \eqref{eq:gluonoperator} by $\lambda_{Xgg}=\lambda_{XQQ} m_Q/v$.
The cross-section contours are in units of the leading-order SM di-Higgs cross section, $\sigma_{hh}^\textrm{SM,LO} = 14.5\,\mathrm{fb}$, 
so by assuming that the scalar $X$ decays into Higgs pairs with a maximal branching fraction, $\Br(X\rightarrow hh)=1$, 
they can be interpreted as the \textit{maximal enhancement} of the LHC di-Higgs cross section over the SM expectations that can be obtained within these models.}
\label{fig:simplifiedmodelxsec}
\end{figure}

Let us first discuss the di-Higgs production rates at the LHC, assuming $\sigma(hh) \sim \sigma_X$.
In Fig.~\ref{fig:simplifiedmodelxsec}, we show contours of $\sigma_X$ at the 13 TeV LHC, normalized to the leading order SM di-Higgs cross section, 
$\sigma_X/\sigma_{hh}^\textrm{SM,LO}$, with $\sigma_{hh}^\textrm{SM,LO}=14.5 \fb$. 
In the three panels of the figure, going from the largest rates to the smallest, 
we assume down quark-fusion production via the operator Eq.~\ref{eq:quarkcoupling} (left), 
1-loop gluon-fusion production via the top-Yukawa operator Eq.~\ref{eq:topcoupling} (middle), 
and gluon-fusion production via the non-renormalizable operator Eq.~\ref{eq:gluonoperator} (right). 
For the case of quark fusion we have taken as an example a scalar coupling to down quarks, 
since in this case large resonant production rates are expected due to the large first-generation quark PDFs.
In fact, we see that di-Higgs rates well above the Standard Model expectations can potentially be obtained in models where scalars are directly produced via this channel.
As an example, cross-sections as large as $10^3$ times the SM di-Higgs cross-section are found for $\mathcal{O}(0.1)$ couplings to down quarks. 
Di-Higgs rates in models with gluon-initiated production may also be larger than the SM expectations, 
but the enhancements are obviously less pronounced, 
as in this case scalars are resonantly produced only at loop level.

Current LHC bounds on resonant di-Higgs production from ATLAS \cite{Aad:2019uzh} and CMS \cite{Sirunyan:2018ayu} are at the level of $\sigma \lesssim 70 \times \sigma_{hh}^{\textrm{SM,LO}} $ for a $300 \textrm{ GeV}$ resonance, 
and improve to $\sigma \lesssim 1 \times \sigma_{hh}^{\textrm{SM,LO}} $ for a $2 \, \textrm{TeV}$ resonance. 
By comparing these bounds with the results from Fig.~\ref{fig:simplifiedmodelxsec}, 
we see that the di-Higgs topology is in principle an efficient tool to test extended Higgs sectors. 
This is especially true in theories where scalars couple to first-generation quarks, 
where percent-level couplings can potentially be probed.
On the other hand, 
models with gluon-initiated production can be probed only if the couplings to the top or gluons are $\mathcal{O}(1)$,
due to the loop suppression in the scalar production.

An honest evaluation of the discovery potential of the di-Higgs topology must be carried within UV complete models, 
both to compute the actual values of $\br(X\rightarrow h h)$, 
and to consider other possible bounds on the model which may be more relevant than di-Higgs production. 
However, it turns out that our simple estimates of the maximal di-Higgs rates in Fig.~\ref{fig:simplifiedmodelxsec} are already a reasonable estimate of the maximal rates within realistic models.
For instance, in the singlet-extension of the SM --- the simplest example of a model with one-loop top-quark mediated production --- it was found in \cite{Chen:2014ask} that di-Higgs rates as large as $\sigma_{hh} \sim 20 \times \sigma_{hh}^{\textrm{SM,LO}}$ can be obtained for $m_X \simeq 300\,\textrm{GeV}$.
Regarding models where scalars are produced via a non-renormalizable contact interaction to gluons,
in \cite{Nakamura:2017irk} it was found that in models where this operator is generated by heavy vector-like quarks, 
di-Higgs rates of order $\sigma(hh)/\sigma_{hh}^\textrm{SM,LO} \sim \mathcal{O}(10)$ can be obtained.
The largest resonant di-Higgs cross sections, where scalars are produced from quark-fusion, have not previously been explored, even though they naively represent the most accessible target, as viable models have only recently been found. 
We discuss these models in detail in later sections.

\begin{figure}[t!]
\vspace{0.5cm}
\quad
\includegraphics[scale=0.4]{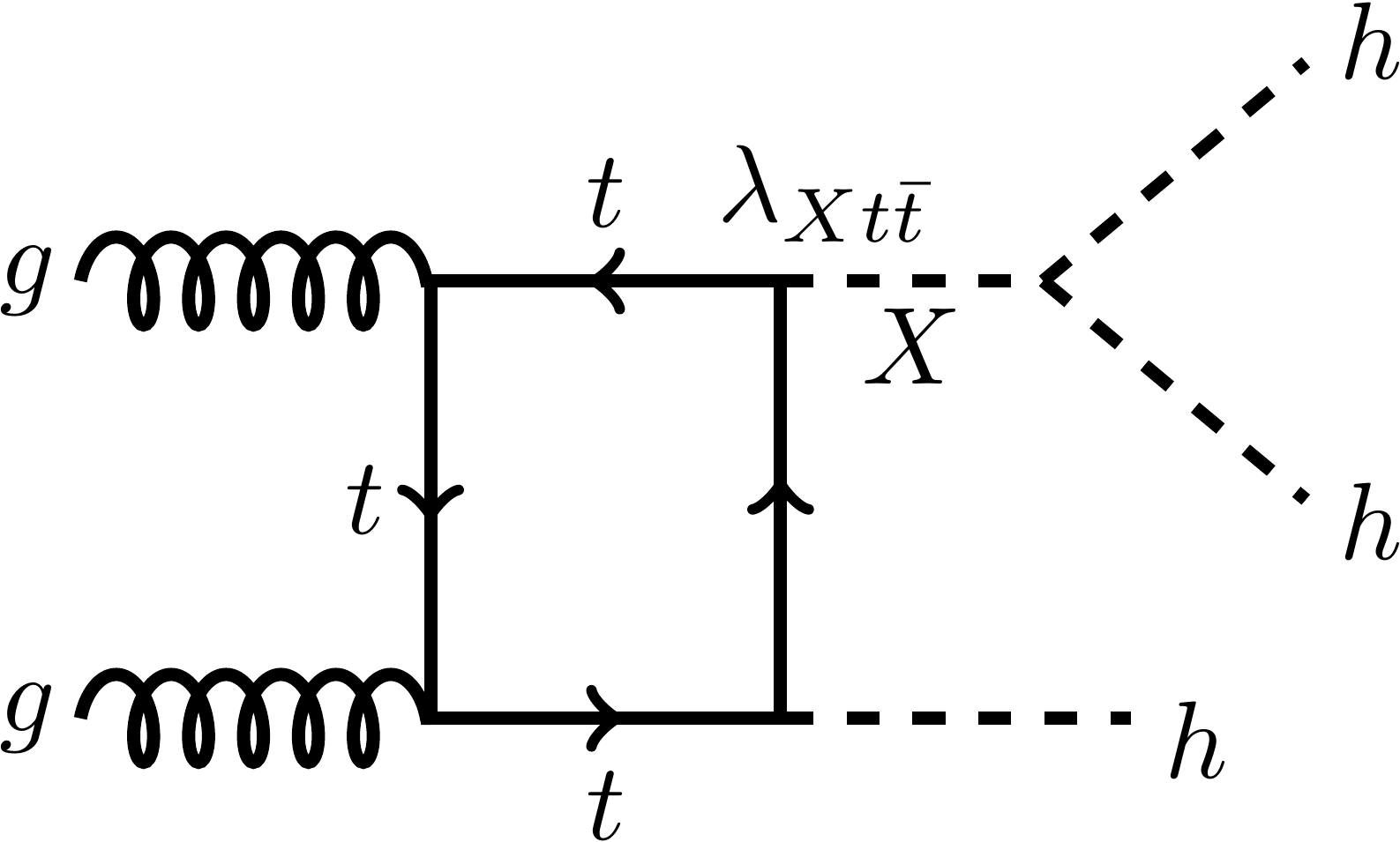}
\hspace{1cm}
\includegraphics[scale=0.4]{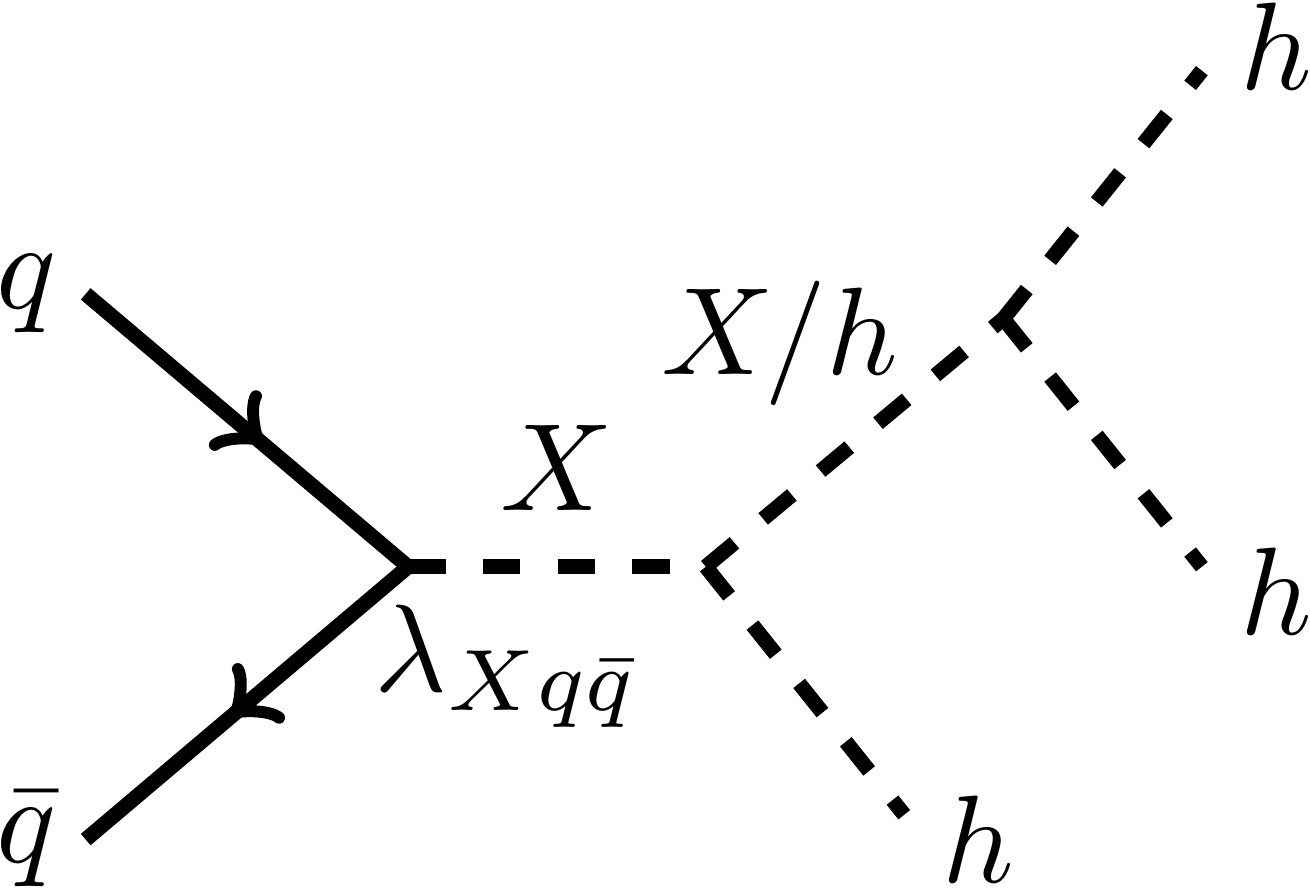}
\caption{Semi-resonant (left) and other resonant and non-resonant (right) tri-Higgs production diagrams.}
\label{fig:diagrams2}
\end{figure}

We now move on to triple-Higgs production. 
From Fig.~\ref{fig:simplifiedmodelxsec}, and assuming a resonant tri-Higgs cross section $\sigma(hhh) \sim 10^{-2}\, \sigma_X$ as discussed above, 
we see that sizable rates for tri-Higgs production can be obtained at the LHC, 
most evidently in models with quark fusion. 
For instance, for $\lambda_{Xd\bar{d}}=0.01$ (which is consistent with  bounds in specific UV completions to be presented later in this work coming from di-Higgs searches) and $m_X=1\, \textrm{TeV}$,  
cross sections as large as $\sim 0.1\, \textrm{fb}$ can be obtained, 
which already lead to a couple of events at the LHC, 
and can be much more precisely probed at the HL-LHC, especially due to the unique characteristics of the final state.
Interestingly, given the couplings discussed above that lead to resonant tri-Higgs production, non-resonant topologies can be automatically realized. 
For instance, semi-resonant $3h$ production can be obtained via associated $hX$ production with $X\rightarrow hh$, 
while fully non-resonant topologies can be realized via off-shell $h/X$ production, as shown in Fig.~\ref{fig:diagrams2}. 
Thus, when studying triple-Higgs production a variety of kinematical features can be realized,
and future studies need to account for the different possibilities in order to properly assess backgrounds.

We conclude by commenting on the case where the mass of the extra scalar states are beyond the reach of direct searches.
In this case, the effects in the low-energy multi-Higgs phenomenology can be described by the SM EFT in the form of non-renormalizable operators involving the Higgs doublet $H_1$.
Integrating out extra scalars with the couplings in Eqs.~\eqref{eq:quarkcoupling}\,--\,\eqref{eq:quarticdecay} leads to dimension-six operators such as
\begin{equation}
(H_1^\dagger H_1)^3 , \qquad  (H_1^\dagger H_1) (Q H_1^{(c)} \bar{q}) ,
\label{eq:SMEFT}
\end{equation}
where $Q$ and $\bar{q}$ are left and right-handed quarks respectively.
The above operators may enhance the non-resonant di-Higgs cross section either by modifying the $125\,\textrm{GeV}$ Higgs trilinear self-coupling in the case of $(H_1^\dagger H_1)^3$, 
or by directly contributing to di-Higgs production via quark-fusion in the case of the cubic-Higgs Yukawa operator $(H_1^\dagger H_1) (Q H_1^{(c)} \bar{q})$~\cite{Egana-Ugrinovic:2019dqu,Alasfar:2019pmn}.
However, 
when the extra Higgses are beyond the reach of resonant production at the LHC,
the effects of these operators are generically too small to be detectable if they arise from perturbative UV completions.
For instance, within the 2HDM the coefficient of the $(H_1^\dagger H_1)^3$ operator is $ 3\lambda_6^2/m_H^2$, where $\lambda_6$ is a quartic coupling of the Higgs potential~\cite{Egana-Ugrinovic:2015vgy}. 
For $m_H=3 \, \textrm{TeV}$, and taking for concreteness $\lambda_6 \lesssim 2$ to avoid Landau poles close to the TeV scale,
the $(H_1^\dagger H_1)^3$ operator leads to an enhancement of the di-Higgs cross section over the SM expectations of order $\mathcal{O}(10\%)$, 
which is too small to be detected even at the HL-LHC~\cite{DiMicco:2019ngk}. 
We will see later that the conclusion is the same for the cubic Higgs Yukawa operator: 
while avoiding Landau poles, retaining consistency with current Higgs boson coupling measurements and bounds from resonant production, 
not even the HL-LHC is sensitive to the di-Higgs rates obtained via this operator.

\subsection{Flavor and SM Higgs couplings}
\label{sec:smhiggscouplings}

An irreducible feature of extended Higgs sectors leading to BSM multi-Higgs production, is that mixing between extra scalars and the Higgs is realized. 
This is easy to see by noting that the operators of Eqs.~\eqref{eq:xicoupling} and \eqref{eq:quarticdecay} are always accompanied by similar ones where a Higgs is replaced by a condensate insertion, 
\begin{equation}
2 \xi X h v , \qquad  3 \lambda X h v^2 ,
\end{equation}
so that mass-mixing between $X$ and $h$ is obtained.
Such mixing leads to modifications to the Higgs couplings with respect to the SM expectations. 
Importantly, given that large multi-Higgs rates are realized at the LHC in models where the extra scalars couple to quarks, 
such rates are then naturally correlated with modifications to the flavored SM Higgs couplings. 
It is then clear that multi-Higgs production is intimately connected with the flavor properties of the Higgs. 
We will now explore this connection in further detail, 
by studying multi-Higgs production in concrete models of extended Higgs sectors.

\section{A UV complete theory of scalars with large couplings to down quarks}
\label{sec:2HDM}

In the previous section we concluded that models where extra scalars couple to light quarks can be efficiently probed via multiple-Higgs production. 
We now present a UV complete theory where such couplings arise naturally. 
This model is treated in detail in \cite{Egana-Ugrinovic:2019dqu}, so here we simply summarize the notation and explain its main features.
For concreteness, we present a theory with large couplings to the down-type quarks, but similar models can be constructed where large couplings to up-type quarks are realized. 
We consider a two-Higgs doublet model with renormalizable Lagrangian given by
\begin{equation}
 D_{\mu}H_{a}^{\dagger}D^{\mu}H_a  - V(H_1, H_2) 
 - \bigg[\lambda^{u}_{aij}Q_i H_a \bar{u}_j 
- \lambda^{d\dagger}_{aij} Q_i {H_a}^c \bar{d}_j 
- \lambda^{\ell \dagger}_{aij} L_i {H_a}^c \bar{\ell}_j 
+ \mathrm{h.c.} \bigg],
\label{eq:2HDMLagrangian}
\end{equation}
where the potential is
\begin{eqnarray}
 \nonumber
 V(H_1,H_2) & = &  m_1^2  H_1^\dagger  H_1+ m_2^2  H_2^\dagger  H_2+\Big( m_{12}^2  H_1^\dagger  H_2 +\textrm{h.c.}\Big) 
 \\
 \nonumber
 &+ &
\frac{1}{2} {\lambda}_1 ( H_1^\dagger  H_1)^2+ \frac{1}{2} {\lambda}_2( H_2^\dagger  H_2)^2 
+{\lambda}_3( H_2^\dagger  H_2)( H_1^\dagger  H_1)+ {\lambda}_4 ( H_2^\dagger  H_1)( H_1^\dagger  H_2) 
\\
& + & 
\bigg[ ~ \frac{1}{2}{\lambda}_5( H_1^\dagger  H_2)^2+{\lambda}_6  H_1^\dagger  H_1  H_1^\dagger  H_2 +{\lambda}_7 ( H_2^\dagger  H_2)( H_1^\dagger  H_2)+\textrm{h.c.}~\bigg] .
\label{eq:2HDMpotential}
\end{eqnarray}
We work in the CP conserving limit, so all the couplings in the potential \eqref{eq:2HDMpotential} can be taken as real. In addition, and without loss of generality, we work in the Higgs basis~\cite{Georgi:1978ri, Botella:1994cs}, where $H_1$ is the SM-like doublet that condenses and gives mass to fermions and gauge bosons, 
while the second doublet does not participate in electroweak symmetry breaking,
\begin{equation}
\langle H_1^{\dagger} H_1 \rangle = \frac{v^2}{2}, \qquad \langle H_2^{\dagger} H_2 \rangle = 0,
\label{eq:higgs_basis}
\end{equation}
where $v = 246\gev$. 

The couplings of the two doublets to the SM fermions are set by their Yukawa matrices in Eq.~\eqref{eq:2HDMLagrangian}. 
In a given flavor basis, the Yukawa matrices of the first doublet are set by the SM Higgs boson Yukawas and the CKM matrix, $V$. 
In what follows we commit to the flavor basis where the first-doublet Yukawa matrices are given by
\begin{eqnarray}
\lambda^u_{1} &=&
V^T Y^u  \equiv V^T  \, \mathrm{diag}(y^{\textrm{SM}}_u, y^{\textrm{SM}}_c, y^{\textrm{SM}}_t) ,
\nonumber \\
\lambda^d_{1} &=& Y^d
 \equiv   \mathrm{diag}(y^{\textrm{SM}}_d, y^{\textrm{SM}}_s, y^{\textrm{SM}}_b) .
\end{eqnarray}
where $y_q^{\textrm{SM}}$ are the SM Higgs boson Yukawas to the different quarks.

The second doublet Yukawa matrices, on the other hand, are free parameters of the model.
To allow for large couplings to light quarks, while avoiding tree-level FCNCs, we require these matrices to be flavor-aligned with the first-doublet Yukawas. 
Flavor-alignment cannot be imposed by symmetries that are manifest in the infrared, but in \cite{Egana-Ugrinovic:2018znw, Egana-Ugrinovic:2019dqu} it was shown that it can be imposed as a UV boundary condition in technically-natural theories called Spontaneous Flavor Violating (SFV). 
Here we focus on up-type SFV theories, where large couplings to the down-type quarks are allowed. 
In up-type SFV, the Yukawa matrices of the second-doublet to quarks are given by
\begin{eqnarray}
\lambda^u_{2} &=&
 \xi \, V^T Y^u ,
\nonumber \\
\lambda^d_{2} &=& K^d
 \equiv  \mathrm{diag}( \kappa_d , \kappa_s,\, \kappa_b ) ,
 \label{eq:uptypeSFV}
\end{eqnarray}
where $\xi $ is a proportionality constant between the first and second doublet up-type Yukawa matrices, and $K^d$ is a real-diagonal matrix. 
Importantly, note that in up-type SFV, the couplings of the second doublet to the down-type quarks are controlled by three free parameters $\kappa_{d,s,b}$, 
which may have arbitrary hierarchies.
Also, note that these couplings are flavor-aligned with the couplings of the first doublet, so they do not lead to tree-level FCNCs.
The couplings to up-type quarks, on the other hand, are proportional to the first-doublet couplings,
so they are also flavor aligned, 
but they maintain the SM flavor hierarchies.
This is a limitation of up-type SFV: the UV completion ensuring alignment only allows for novel flavor hierarchies in the down-quark sector.\footnote{Another prescription, called down-type SFV allows for novel hierarchies in the up-sector instead.}

For simplicity, in what follows we make a set of assumptions regarding couplings to fermions that allow us to focus on the phenomenology of new Higgs states with large couplings to light quarks.
First, we only consider non-zero $\kappa_d$ or $\kappa_s$, and set $\kappa_b=0$. 
Second, we suppress couplings to up-type quarks by setting $\xi=0$.
Finally, we make our second-doublet leptophobic by setting its lepton Yukawa matrix to zero, $\lambda^{\ell}_{2ij}=0.$

The physical Higgs bosons of the second doublet and their couplings are most easily obtained in the unitary gauge. 
In this gauge, the first Higgs doublet $H_1$ contains one CP even neutral state $h_1$. The second Higgs doublet $H_2$ contains one CP even state $h_2$, one CP odd neutral state $A$, and one new charged Higgs boson $H^\pm$. The latter two states are mass eigenstates, but $h_1$ and $h_2$ mix in the mass matrix, giving rise to the $125\,\textrm{GeV}$ SM Higgs boson $h$, and a new CP even Higgs state $H$ with mass $m_H$. 
The mixing is controlled by the alignment parameter, $\cos(\beta-\alpha)$, and is given by
\begin{eqnarray}
\begin{split}
\label{eq:higgs_mass_eigenstates}
h & \equiv \sin(\beta - \alpha) h_1 + \cos(\beta - \alpha)h_2 , \\
H & \equiv -\cos(\beta - \alpha) h_1 + \sin(\beta - \alpha) h_2 , \\
\end{split}
\end{eqnarray}
where the alignment parameter in terms of the Higgs potential couplings of Eq.~\eqref{eq:2HDMpotential} is 
\begin{equation}
\tan\big[ 2(\beta - \alpha) \big] = \frac{2 \lambda_6 v^2}{\lambda_1 v^2 - \big( m_2^2 +  \frac{1}{2}(\lambda_3 + \lambda_4 + \lambda_5)v^2 \big)} .
\label{eq:alignment_angle}
\end{equation}
The limit $\cos(\beta-\alpha)\rightarrow 0$ is called the \textit{alignment limit}, and is obtained by setting $\lambda_6=0$ in the Higgs potential, 
or by going to the decoupling limit $m_H\rightarrow \infty$. 
In the alignment limit,  
the couplings of $h$ to SM fermions and gauge bosons are SM-like. 
Non-zero alignment parameter signals a deviation from the SM predictions for the Higgs couplings, 
which can be bound by current measurements~\cite{Craig:2013hca}.

In terms of the alignment parameter, the couplings of the Higgs mass eigenstates to fermions are summarized in Table~\ref{t:yukawaup}. The couplings of the neutral states to two gauge bosons $V=W,Z$ are given by
\begin{eqnarray}
\nonumber \lambda_{HVV} &=& -\frac{2m_V^2}{v} \cos(\beta-\alpha) , \\
\lambda_{AVV} &=& 0 .
\label{eq:gaugebosoncoupling}
\end{eqnarray}
The rest of the couplings to gauge bosons are given in \cite{Gunion:2002zf}.
The couplings of the new Higgs mass eigenstates to the SM Higgs are obtained by using the mixing matrix Eq.~\eqref{eq:higgs_mass_eigenstates} in the Higgs potential Eq.~\eqref{eq:2HDMpotential}.
The exact expressions for such couplings are lengthy and can be found in \cite{Gunion:2002zf}. However, simplified and illustrative expressions can be obtained near the decoupling limit, where $m_H \gg m_h$. 
In this case, the alignment parameter is controlled by a single quartic coupling of the Higgs potential, $\lambda_6$, and it is given by, 
\begin{equation}
\cos(\beta-\alpha)=-\lambda_6 \frac{v^2}{m_H^2} + \mathcal{O}\left(\frac{v^4}{m_H^4}\right).
\label{eq:approxalign}
\end{equation}
In the decoupling limit, the couplings of the heavy Higgs bosons to two and three SM bosons are controlled by the same quartic coupling, and can be approximated by
\begin{eqnarray}
 \frac{1}{v}\lambda_{Hhh}= &=& -3 \lambda_6  + \mathcal{O}\left(\frac{v^2}{m_H^2}\right)  
=  \frac{m_H^2}{v^2} \cos(\beta-\alpha) + \mathcal{O}\left(\frac{v^2}{m_H^2}\right), 
\label{eq:dihiggscoupling}
 \\
 \lambda_{Hhhh} &=& -3 \lambda_6  + \mathcal{O}\left(\frac{v^2}{m_H^2}\right)  
=  \frac{m_H^2}{v^2} \cos(\beta-\alpha) + \mathcal{O}\left(\frac{v^2}{m_H^2}\right) , 
\label{eq:dihiggscoupling2}
\end{eqnarray}
where we use the couplings conventions of \cite{Egana-Ugrinovic:2015vgy}.
The pseudoscalar Higgs boson $A$ does not couple to two or three SM Higgs bosons in our CP conserving theory. 
This means in particular that resonant di- or tri-Higgs production happens only via the new scalar Higgs $H$.
Note that in the alignment limit obtained by setting $\lambda_6=0$, the coupling of the new scalar $H$ to two or three SM Higgses vanishes. From Eqs.~\eqref{eq:dihiggscoupling} and \eqref{eq:dihiggscoupling2}  and Table~\ref{t:yukawaup}, we see the connection between BSM physics in multi-Higgs final states, modifications of the SM Yukawas, and increased multi-Higgs production cross sections. To have new physics leading to multi-Higgs final states we must be {\em away} from the alignment limit, therefore there {\em will} be modifications of the SM Higgs Yukawas inherited from the BSM Higgs.  Therefore large changes in the SM light quark Yukawas are correlated strongly with increased multi-Higgs production cross sections. 

\begin{table}[t!]
\begin{center}
$
\begin{array}{|c|c|c|c|}
\hline
\lambda_{h u_i \bar{u}_j}
& 
 \delta_{ij} Y^u_i \sin(\beta - \alpha) 
&
\lambda_{H u_i \bar{u}_j}
&
- \delta_{ij} Y^u_i \cos(\beta - \alpha) 
\\
\lambda_{h d_i \bar{d}_j} 
&
\delta_{ij} \left[ Y^d_i \sin(\beta - \alpha) + K^d_i \cos(\beta - \alpha) \right]
&
\lambda_{H d_i \bar{d}_j} 
&
 \delta_{ij} \left[ -Y^d_i \cos(\beta - \alpha) + K^d_i \sin(\beta - \alpha) \right]
 \\
 \lambda_{h \ell_i \bar{\ell}_j} 
&
\delta_{ij} Y^\ell_i \sin(\beta - \alpha) 
&
\lambda_{H \ell_i \bar{\ell}_j} 
&
- \delta_{ij}  Y^\ell_i \cos(\beta - \alpha) \\
\lambda_{A u_i \bar{u}_j}
&
0&
\lambda_{H^+ d_i \bar{u}_j}
&
0
\\
\lambda_{A d_i \bar{d}_j}
&
-i
\delta_{ij} 
K^d_i
&
\lambda_{H^- u_i \bar{d}_j}
&
\big[
V^*
K^d
\big]_{ij}
\\
\lambda_{A \ell_i \bar{\ell}_j}
&
0
&
\lambda_{H^- \ell_i \bar{\ell}_j}
&
0
\\
\hline
\end{array}
$
\end{center}
\caption{Higgs bosons couplings to the fermion mass eigenstates with left-handed chirality, 
in a theory with second-doublet Yukawas~\eqref{eq:uptypeSFV},  
with the sign convention $\mathcal{L} \supset - \lambda_{h f \bar{f}} h \bar{f} f$.
$Y^{u,d,\ell}$ are diagonal matrices containing the SM Yukawa couplings to quarks and leptons, 
$V$ is the CKM matrix, and $\cos(\beta-\alpha)$ is the Higgs alignment parameter Eq.~\eqref{eq:alignment_angle}.
In this work and in this table we consider a second doublet that couples only to down-type quarks, 
to simplify the model's phenomenology while allowing for resonant di-Higgs production via quark-fusion. 
For this reason, 
we set the flavor-diagonal down-type quark Yukawa matrix to be down-specific, $K^{d}=\mathrm{diag}( \kappa_d,\, 0,\, 0 )$, 
or strange-specific $K^{d}=\mathrm{diag}( 0,\, \kappa_s,\, 0 )$.
However, note that even when the second-doublet Yukawa matrices for leptons and up-type quarks are set to zero for simplicity,  the extra Higgs states still inherit couplings to the SM leptons and up-type quarks via mixing with the SM Higgs.
Thus, these couplings are proportional to $\cos(\beta - \alpha)$.
}
\label{t:yukawaup}
\end{table}

\section{Double and triple Higgs production in the 2HDM}
\label{sec:dihiggs}

We now evaluate the resonant di- and tri-Higgs production rates at LHC for the theory presented in the previous section.
In this section we mostly concentrate on extra Higgses with large couplings to the down quark, which will allow us to explore quark-fusion multi-Higgs production in detail, 
but we will also comment on tri-Higgs production in the absence of such large couplings.
We postpone the exploration of large couplings to the strange quark to Section~\ref{sec:enhancements}.

\begin{figure}[t]
\includegraphics[width=.45\linewidth]{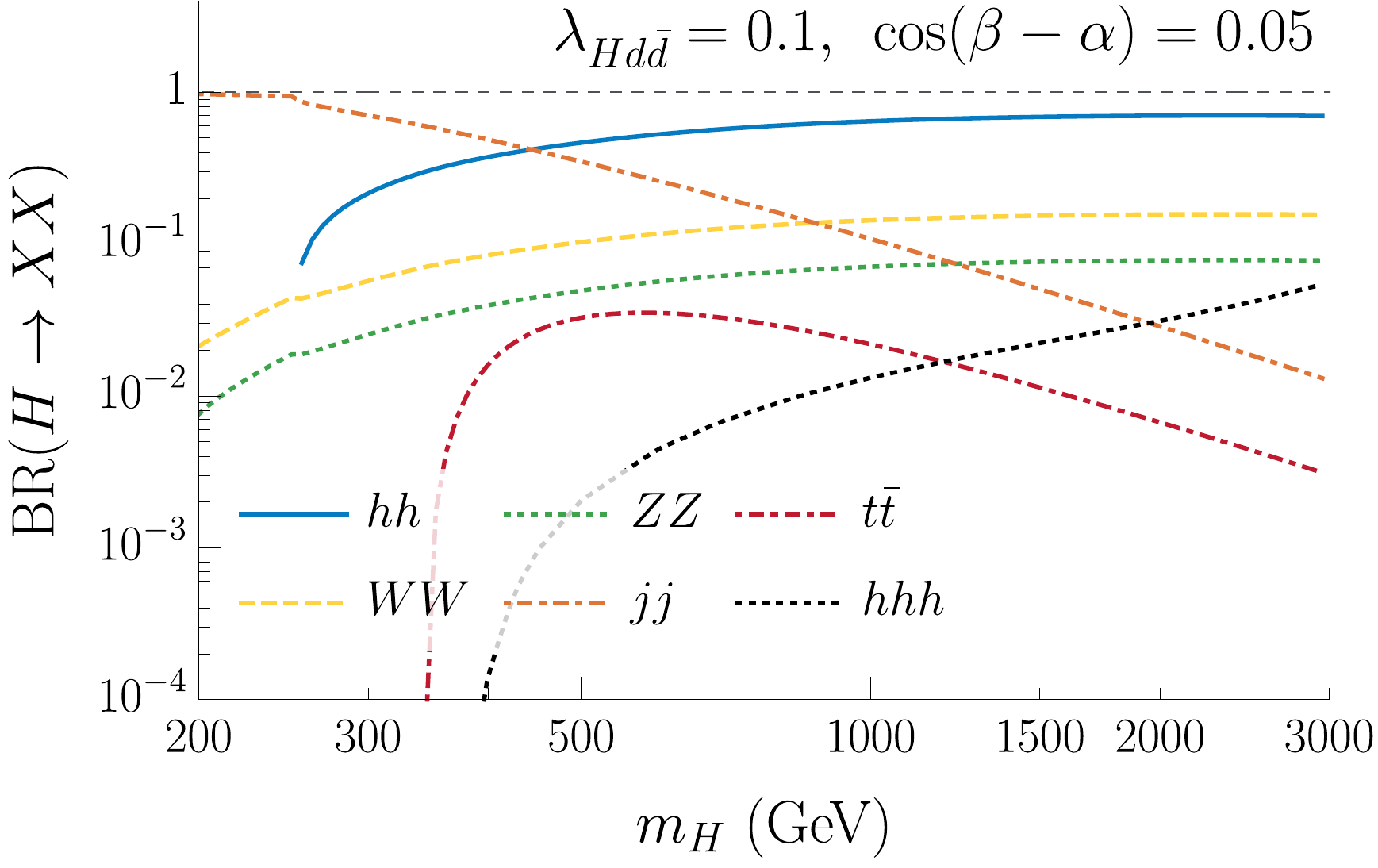}
~
\includegraphics[width=.45\linewidth]{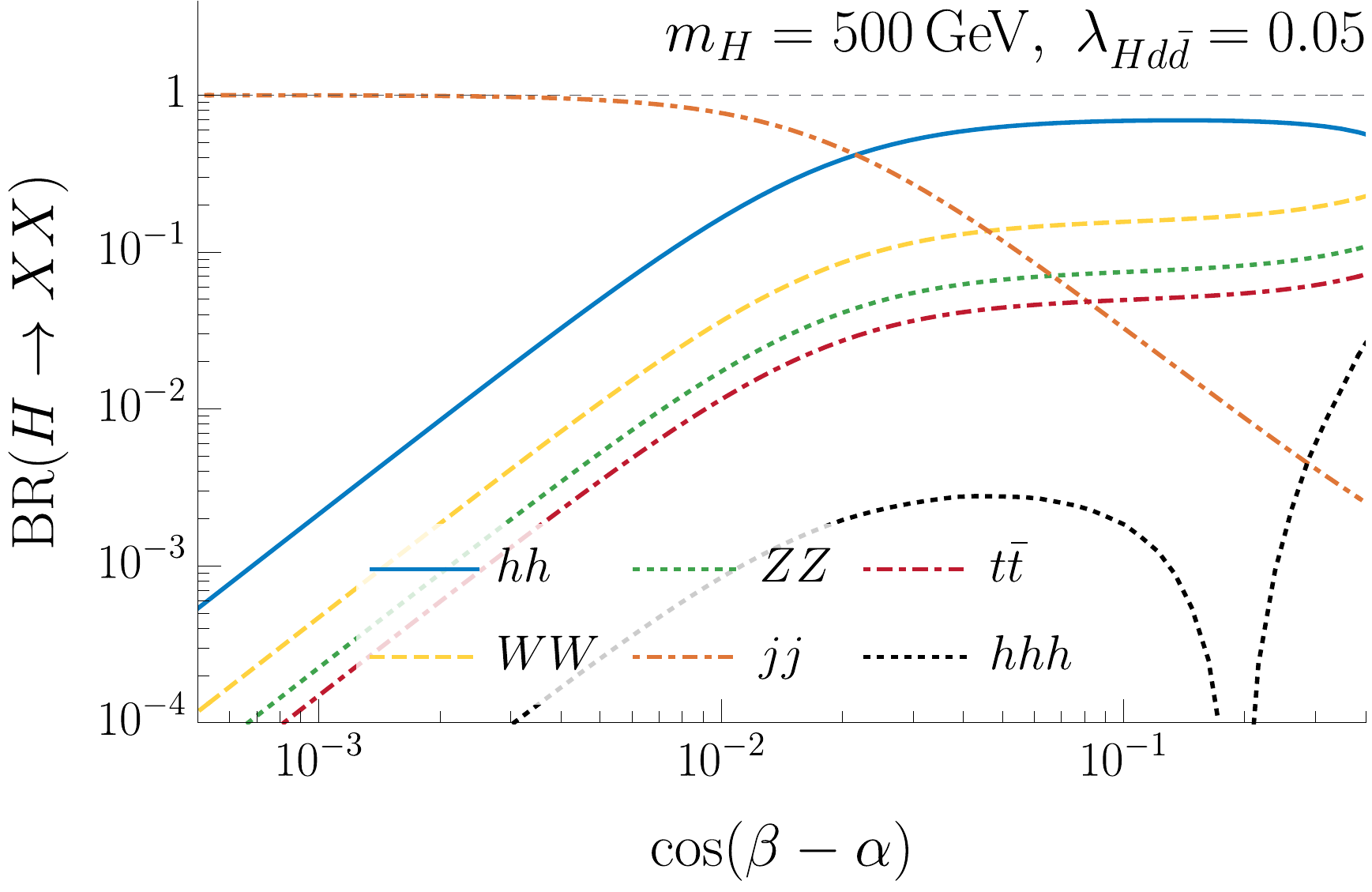}
~
\includegraphics[width=.45\linewidth]{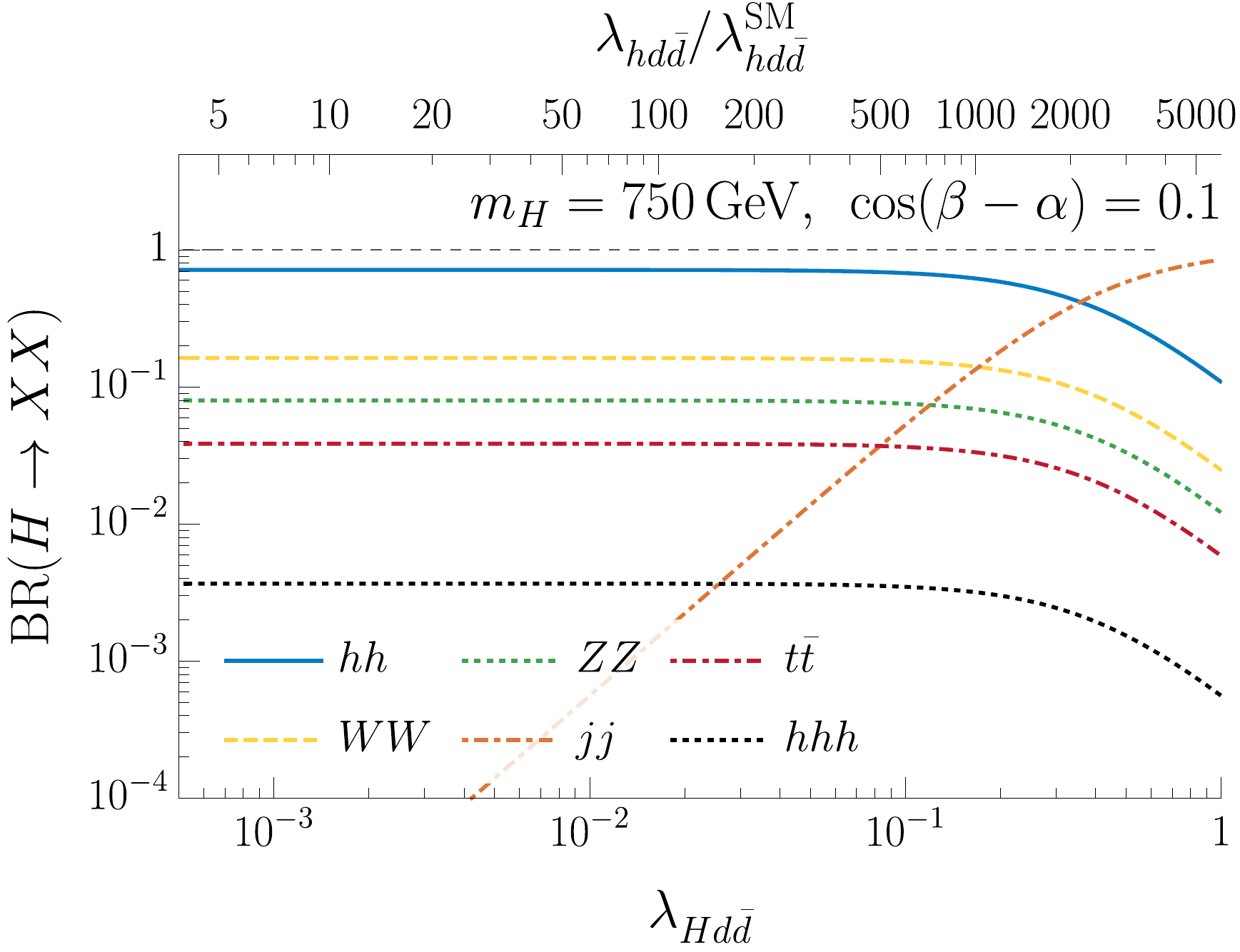}
\caption{
Branching ratios for the the Heavy Higgs $H$ in the up-type SFV 2HDM into various final states as a function of mass (upper-left), alignment parameter  (upper-right), and coupling to down quarks (lower panel).  To emphasize the correlation to shifts in the SM quark Yukawas, in the lower panel we also show the ratio of the physical Yukawa of the 125 GeV Higgs compared to the SM prediction.
}
\label{fig:brs}
\end{figure}

We begin by studying the branching fractions of the extra Higgs $H$ when it can have large couplings to the down quark. In addition to its decays into SM Higgses, $H$ can decay into SM fermions, gauge bosons, and to the extra Higgses $AA$ or $H^+ H^-$.
For simplicity, 
in what follows we choose parameters in the Higgs potential such that all the extra Higgses are degenerate, so the decays to $AA$ and $H^+ H^-$  are kinematically forbidden.\footnote{This is done by setting $\lambda_3=\lambda_4=0,\lambda_5=m_h^2/v^2-\lambda_1$. We also set $\lambda_7=0$ for simplicity.}
With these assumptions, in the three panels of Fig.~\ref{fig:brs}, 
we plot the branching fractions as a function of the $H$ mass $m_H$ (upper left),  the alignment parameter $\cos(\beta-\alpha)$ (upper right), and the coupling of $H$ to down quarks $\lambda_{Hd \bar{d}}$ (lower panel). 
We see that a large branching fraction into two $125\,\textrm{GeV}$ Higgses is obtained over wide regions of parameter space. 
We also observe that the dominant decay mode is always either the decay into two $125\,\textrm{GeV}$ Higgses or decays into down quarks,
while the branching fractions into gauge bosons, the top quark, and leptons are comparatively smaller.  
Decays to gauge bosons are also subdominant since the couplings of $H$ to $125\,\textrm{GeV}$ Higgses are larger than those to gauge bosons by a factor $\lambda_{Hhh}/\lambda_{HVV} \sim m_H^2/m_V^2$, $V=W,Z$ (c.f. Eqs.~\eqref{eq:gaugebosoncoupling} and \eqref{eq:dihiggscoupling}).
The branching fractions into leptons and the top are subdominant since they only arise via mixing with the SM Higgs (as we have set the second-doublet Yukawa to leptons and the top to zero), 
so they are suppressed by one power of the alignment parameter.
Furthermore, couplings to leptons are chirally suppressed by the smallness of the corresponding SM Yukawas.

Now, from the upper-left panel of Fig.~\ref{fig:brs}, we see that for fixed $\cos(\beta-\alpha)$ and $\lambda_{Hd \bar{d}}$, 
decays into two $125\,\textrm{GeV}$ Higgses dominate over decays into down quarks at large $m_H$. 
This is due to an enhancement of the di-Higgs couplings at large masses by a factor $m_H^2/v^2$, c.f Eq.~\eqref{eq:dihiggscoupling}.
However, even for sub-TeV heavy Higgs masses, 
large branching fractions into two Higgses can be obtained.
The only exceptions are below the $hh$ threshold, or when the coupling to the down quark is parametrically larger than the alignment parameter, as can be seen from the lower panel for large $\lambda_{Hd \bar{d}}$.
Note that when the dominant decay mode is into light-quark jets
the collider phenomenology is dominated by multijet searches.
This scenario was investigated in \cite{Egana-Ugrinovic:2019dqu}, 
where this theory in the limit $\cos(\beta-\alpha) \to 0$ was investigated.
In what follows we instead focus on nonzero values of the alignment parameter, for which the multi-Higgs topology may dominate the collider signatures.

One of the most interesting features that we wish to point out here regarding our 2HDM, is that from Fig.~\ref{fig:brs} we see that sizable branching fractions into three Higgs bosons can be obtained, which can be as large as a few percent. 
The largest decay ratios to $3h$ are obtained at large $m_H$, as can be seen from the upper left panel, and when the alignment parameter is large, as seen in the upper right panel. 
In the upper right panel a notable feature arises at $\cos(\beta-\alpha)\sim 0.2$, 
where the branching fraction to $3h$ drops to zero. 
This is due to a cancellation between the direct $H\rightarrow 3h$ amplitude mediated by the coupling in Eq.~\eqref{eq:dihiggscoupling2} and the indirect off-shell processes $H\rightarrow h H^*$ with $H^*\rightarrow hh $ and $H\rightarrow h h^*$ with $h^*\rightarrow hh$.

\begin{figure}[ht]
\includegraphics[width=0.49\linewidth]{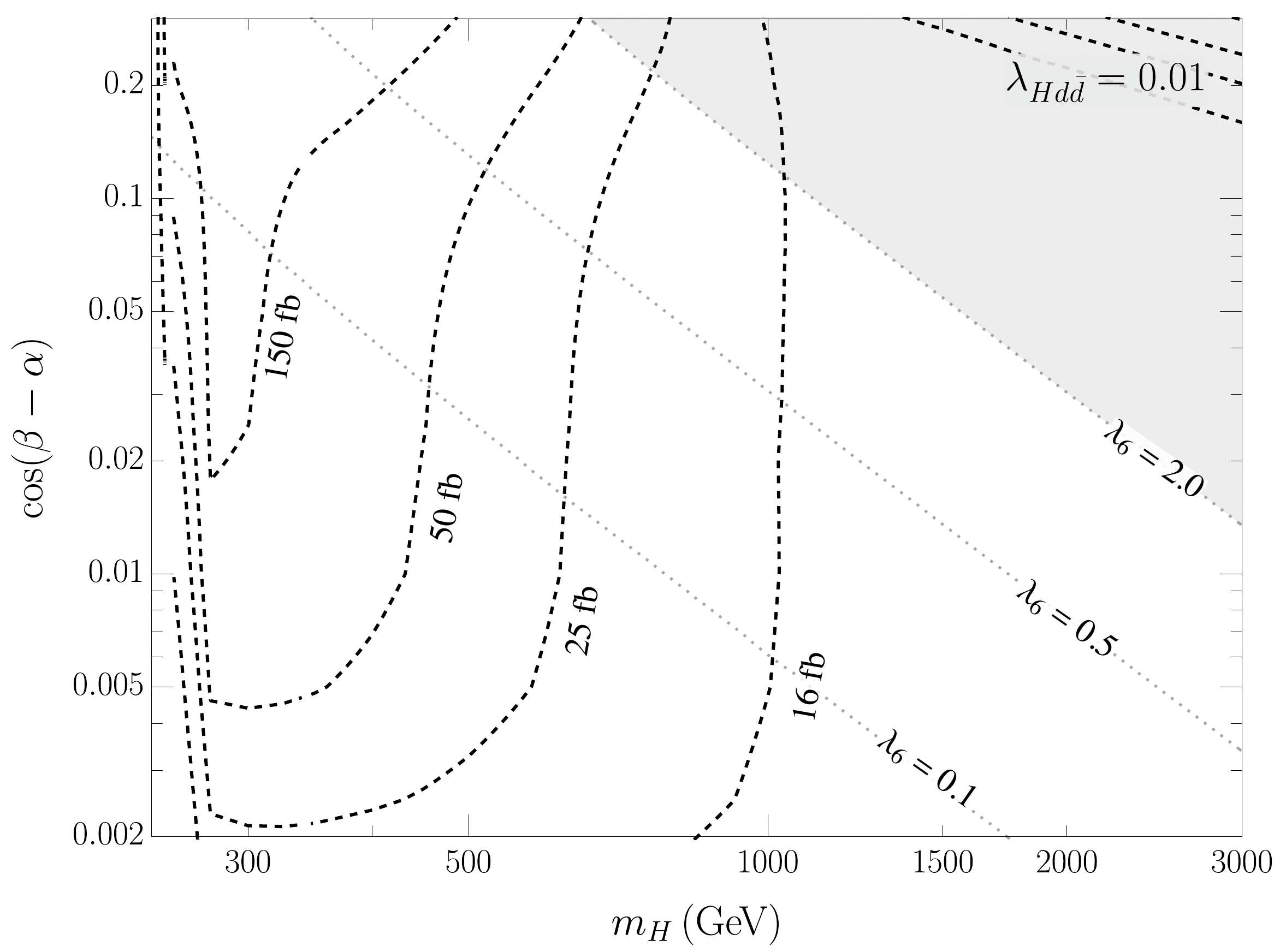}
\includegraphics[width=0.49\linewidth]{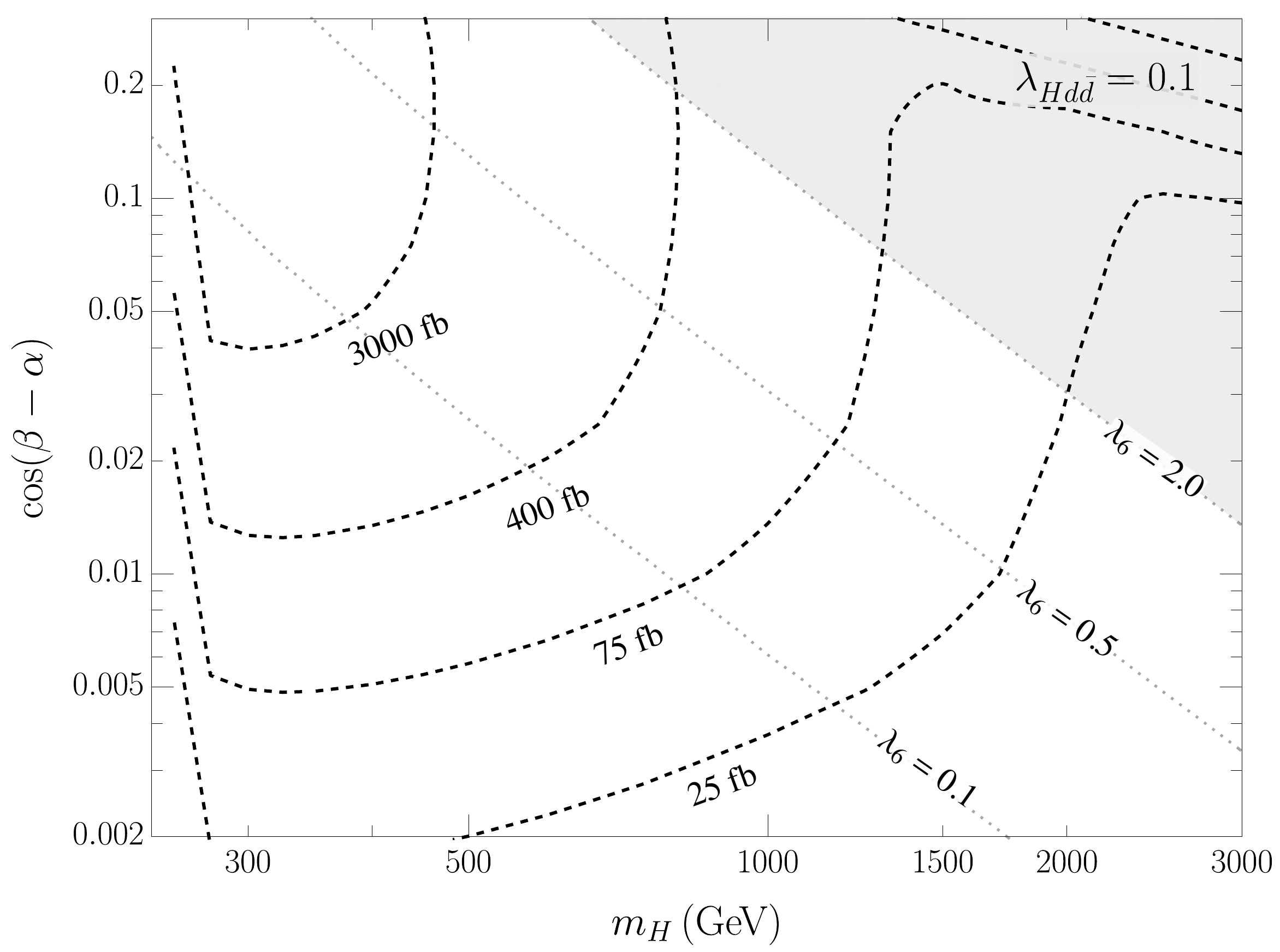}
\includegraphics[width=0.49\linewidth]{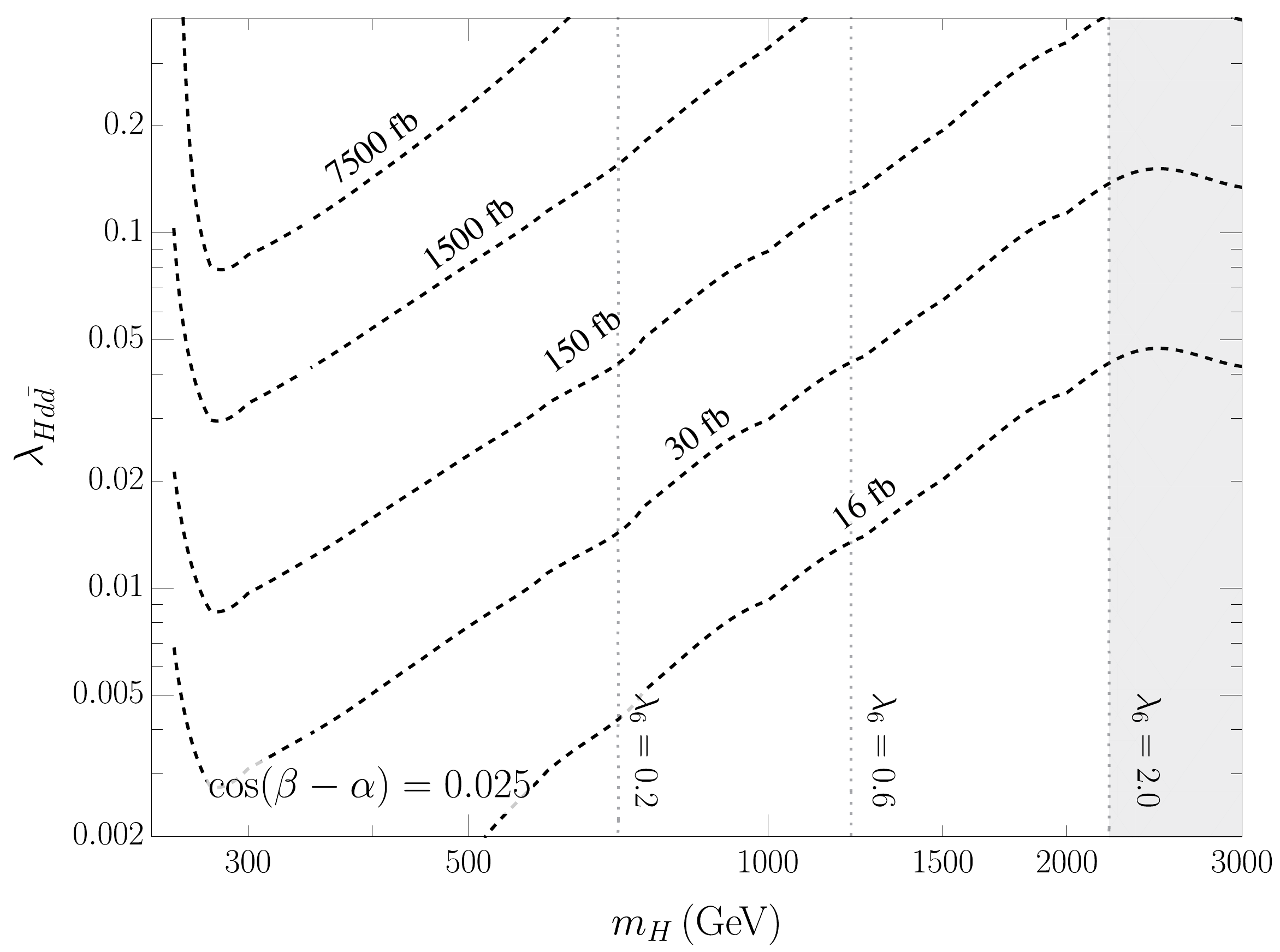}
\includegraphics[width=0.49\linewidth]{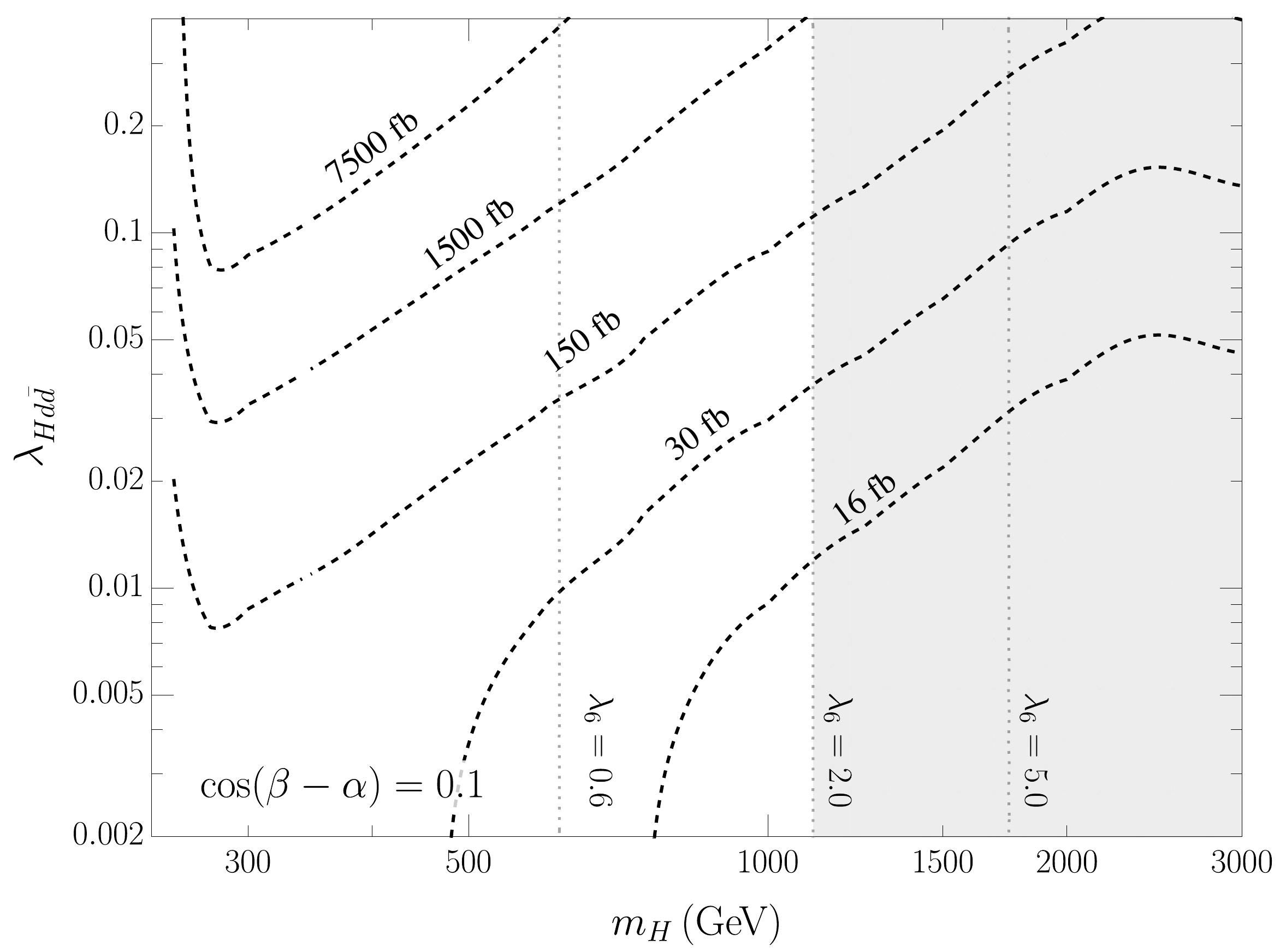}
\caption{
Contours of the di-Higgs production cross section at the 13 TeV LHC for the 2HDM. 
In the top two panels we present the contours as a function of the second scalar Higgs mass $m_H$ and alignment parameter $\cos(\beta-\alpha)$, for fixed $H$ coupling to the down quark $\lambda_{Hd\bar{d}}=0.01$ (top left) and $\lambda_{Hd\bar{d}}=0.1$ (top right).
In the bottom two panels we present the cross section as a function of $m_H$ and $\lambda_{Hd\bar{d}}$ for fixed $\cos(\beta-\alpha)=0.025$ (bottom left) and $\cos(\beta-\alpha)=0.1$ (bottom right).
Rates should be compared to the LO SM di-Higgs cross section at the 13 TeV LHC, $\sigma_{hh}^{\textrm{SM, LO}} = 14.5\,\mathrm{fb}$ (for reference, the NNLO value is $\sigma_{hh}^{\textrm{SM, NNLO}} = 31.05\,\mathrm{fb}$~\cite{Grazzini:2018bsd}).
}\label{fig:hhrates}
\end{figure}

\begin{figure}[ht]
\includegraphics[width=0.49\linewidth]{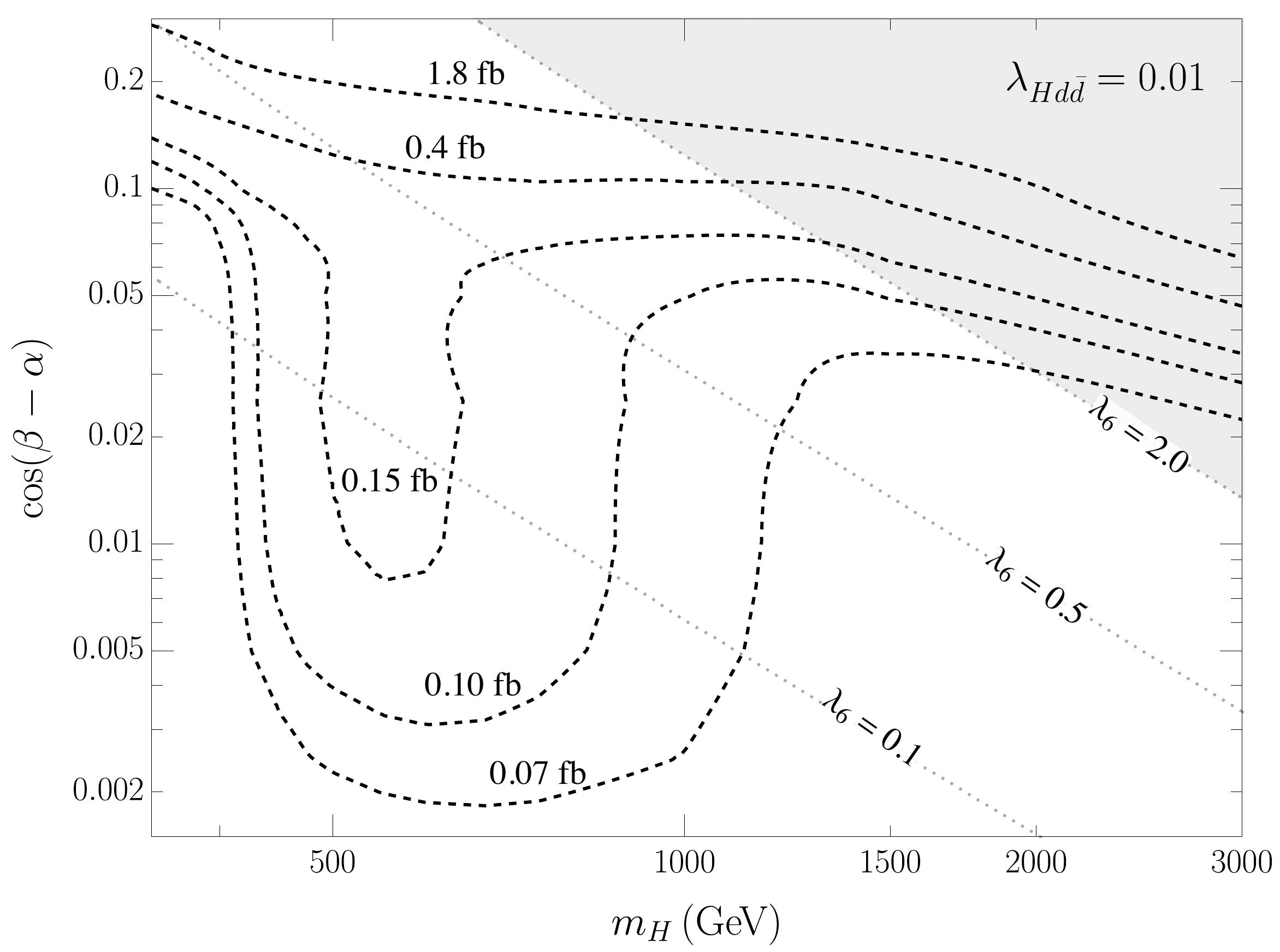}
\includegraphics[width=0.49\linewidth]{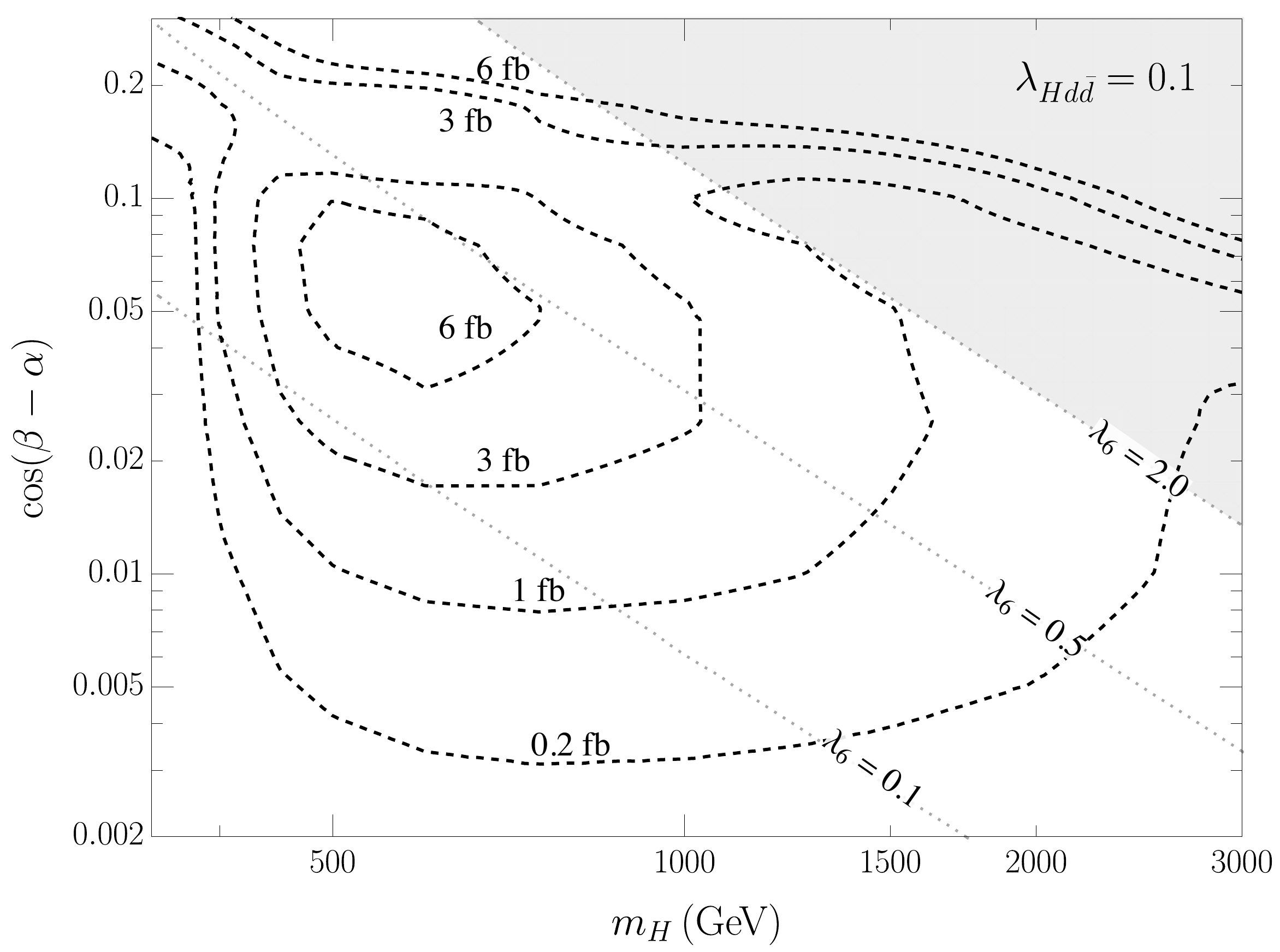}
\includegraphics[width=0.49\linewidth]{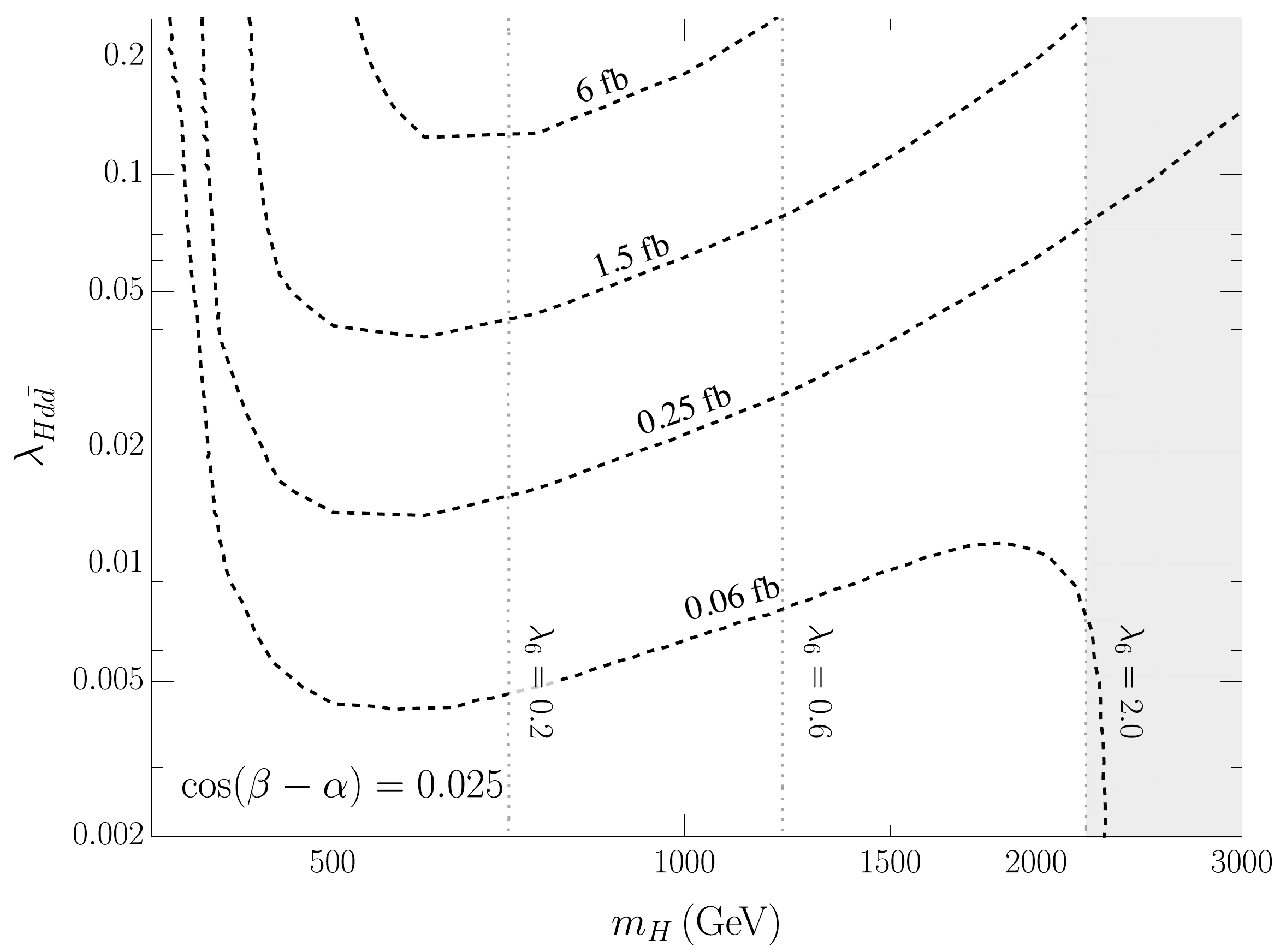}
\includegraphics[width=0.49\linewidth]{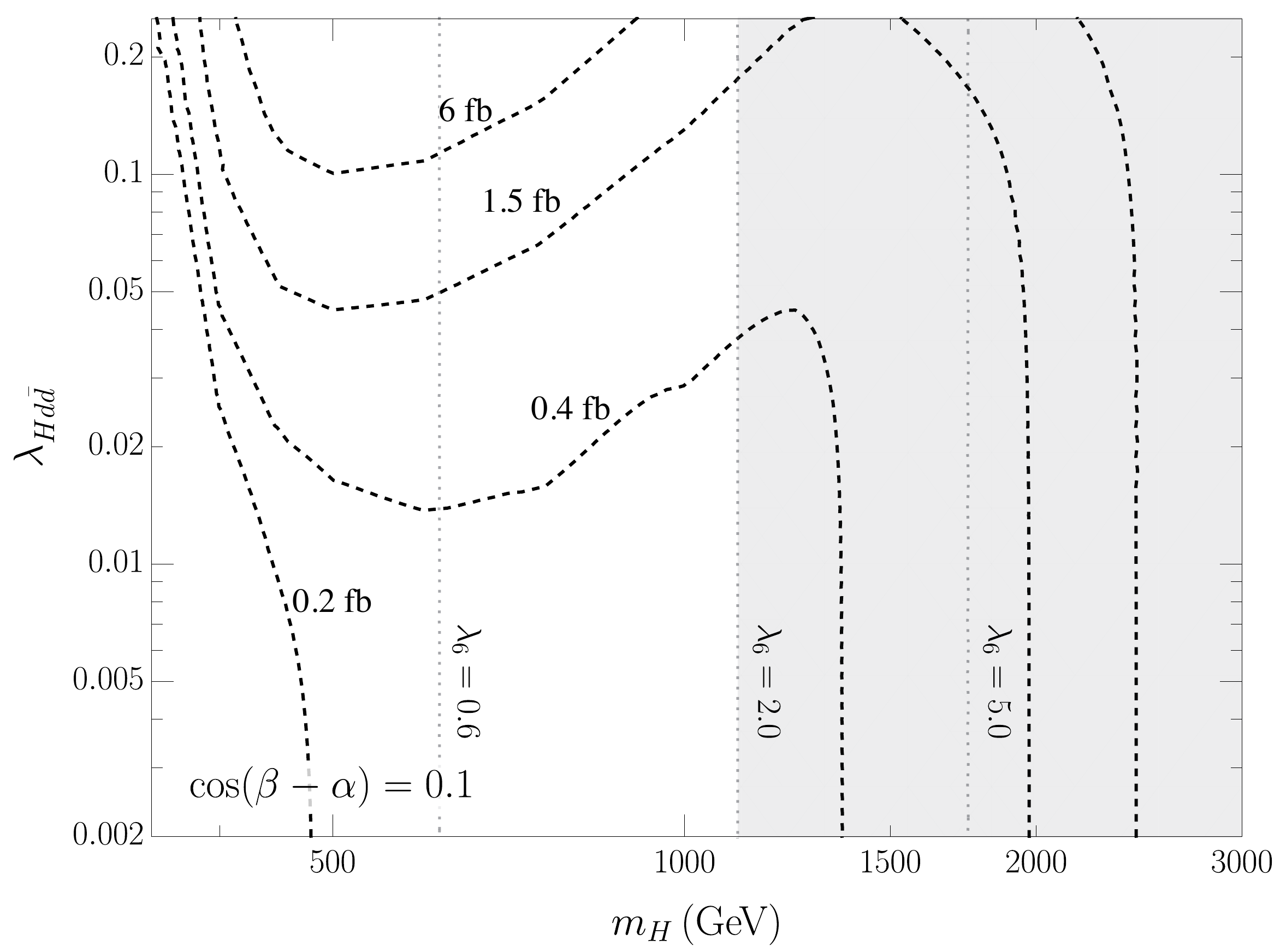}
\caption{
Contours of the tri-Higgs production cross section $\sigma_{hhh}$  at the 13 TeV LHC for the 2HDM, 
as a function of $m_H$ and $\cos(\beta-\alpha)$ for fixed $\lambda_{Hd\bar{d}}=0.01$ or $\lambda_{Hd\bar{d}}=0.1$  (top two panels), or $m_H$ and $\lambda_{Hd\bar{d}}$ for fixed $\cos(\beta-\alpha)=0.025$ or  $\cos(\beta-\alpha)=0.1$ (bottom two panels).
Rates should be compared to the LO SM tri-Higgs cross section at the 14 TeV LHC, $\sigma_{hhh}^{\textrm{SM, LO}} = 0.035\,\mathrm{fb}$ (for reference,  the NNLO value is $\sigma_{hhh}^{\textrm{SM, NNLO}} = 0.103\,\mathrm{fb}$~\cite{deFlorian:2019app}).
}\label{fig:hhhrates}
\end{figure}

Given the above branching fractions, contours of the LO di- and tri-Higgs rates (in fb) at the 13 TeV LHC are presented in Figs.~\ref{fig:hhrates} and \ref{fig:hhhrates}, including the full matrix element for quark and gluon initiated production. 
As a perturbativity requirement, in the figures we have shaded in gray the regions of parameter space where the quartic $\lambda_6$ that controls the alignment parameter (c.f. Eq.~\eqref{eq:approxalign}), 
becomes $\lambda_6\geq 2$, 
since in this case we have checked that Landau poles arise already near the TeV scale.
Consistent with this requirement, 
we see that large rates and huge enhancements over the SM expectations for both di and tri-Higgs production can be obtained in this model, confirming our expectations from the discussion in Section~\ref{sec:simplified}.
In the 2HDM, di-Higgs production happens due to resonant production of $H$ via gluon or quark-fusion, which subsequently decays to two $125\,\textrm{GeV}$ Higgses.

\begin{figure}[ht!]
\includegraphics[width=0.49\linewidth]{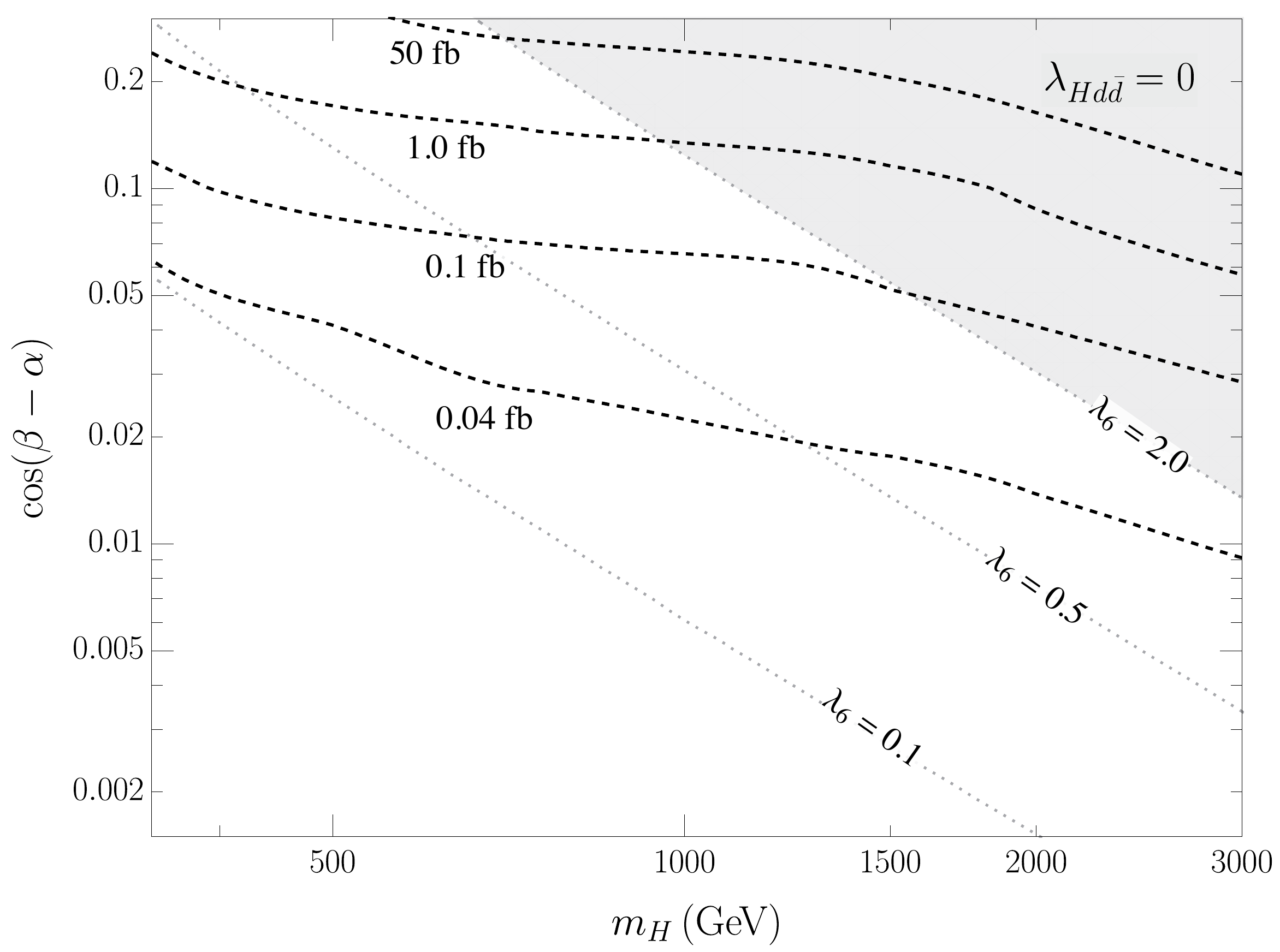}
\caption{
Contours of the tri-Higgs production cross section $\sigma_{hhh}$  at the 13 TeV LHC for the 2HDM, 
as a function of $m_H$ and $\cos(\beta-\alpha)$, in the absence of large couplings to light quarks, $\lambda_{Hd\bar{d}}=\lambda_{Hs\bar{s}}=0$. 
In this case, tri-Higgs production can still arise from top-mediated diagrams as in Fig.~\ref{fig:diagrams2}. 
As we have set the second-doublet Yukawa matrices to zero for simplicity (so this is essentially the case of a type I 2HDM at large $\tan\beta$), here couplings of $H$ to the top arise only due to mixing with the SM Higgs (c.f. Table~\ref{t:yukawaup}), and tri-Higgs production is thus controlled by $\cos(\beta-\alpha)$.}\label{fig:hhhrateskd0}
\end{figure}

Tri-Higgs production, on the other hand, has more complex kinematics, 
as it can arise both from resonant or non-resonant $H$ production with $H \to 3h$ and from gluon-initiated, $hH$ associated production, with $H\rightarrow 2h$, as shown in the diagrams in Fig.~\ref{fig:diagrams2}.
The relevance of these two production modes in different parts of parameter space can be seen by comparing the upper two and lower two panels in Fig.~\ref{fig:hhhrates}.
In the lower panels we see that at large $\lambda_{Hd\bar{d}}$ and small $\cos(\beta-\alpha)$, $\sigma_{hhh}$ has a clear and sharp drop for masses below $m_H\leq3m_h$, as in this case $3h$ production comes mostly from resonant down-quark fusion, $qq\rightarrow H \rightarrow 3h$.
In the upper-left panel, on the other hand, we note that for large $\cos(\beta-\alpha)$ and small $\lambda_{Hd\bar{d}}$, which is a situation that also can happen naturally in e.g., the types I-IV 2HDMs or singlet-extended SM, 
$\sigma_{hhh}$ does not go to zero as the threshold $m_H=3m_h$ is approached. 
In fact, even for $\lambda_{Hd\bar{d}}=0$ (which is essentially the case of a type I 2HDM at large $\tan\beta$), large triple-Higgs rates can be obtained, as shown in Fig.~\ref{fig:hhhrateskd0}. 
The reason is that in this case, tri-Higgs production can still arise, now from $gg\rightarrow hH$ with $H\rightarrow hh$, a process that is controlled by the coupling of $H$ to the top quark.
Thus, within the 2HDM, tri-Higgs production can be used as a probe of extended Higgs sectors with couplings to both light and heavy quarks.


\FloatBarrier
\section{LHC bounds on light quark Yukawas}
\label{sec:enhancements}

\begin{figure}[htp!]
\includegraphics[width=0.66\linewidth]{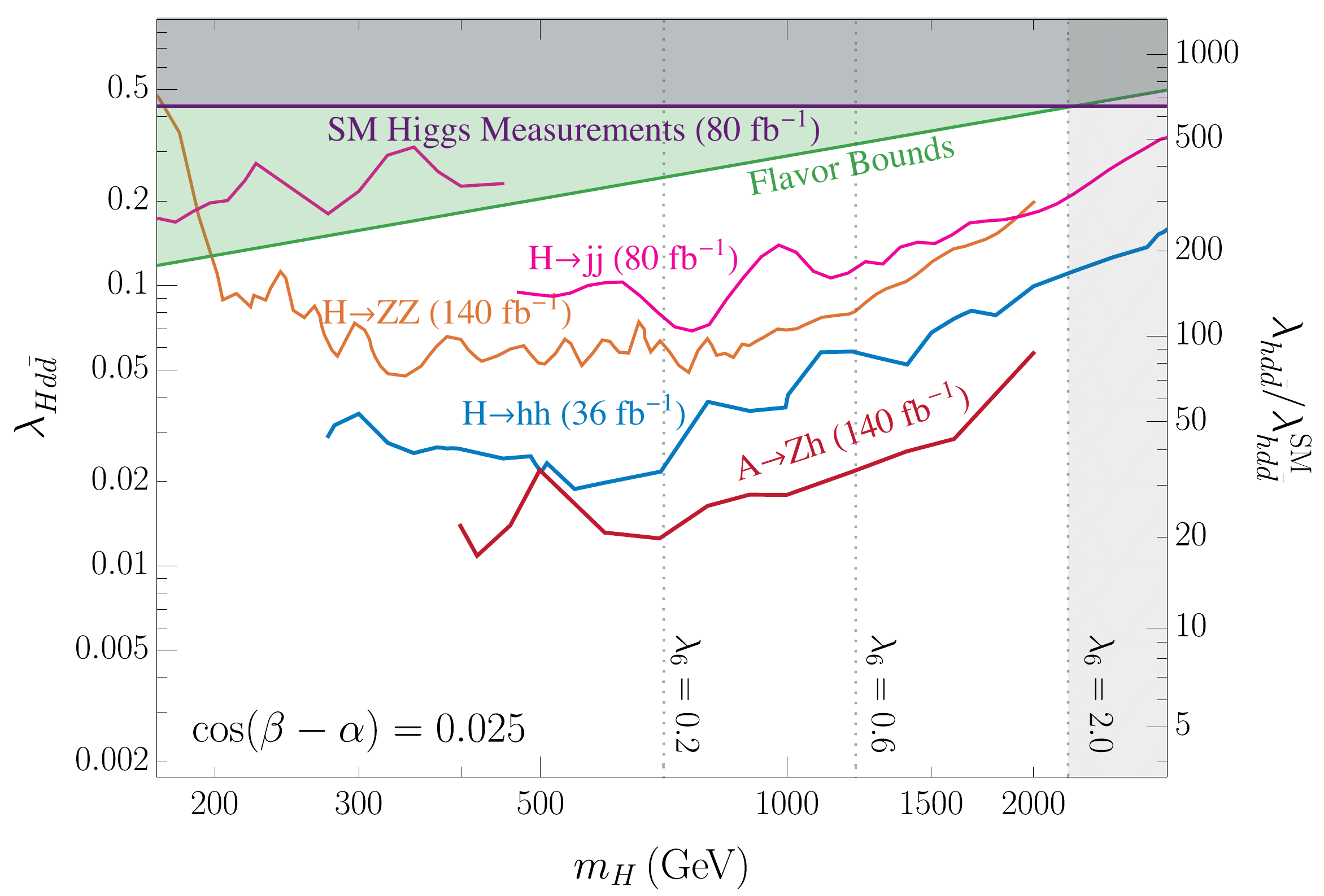}
\vskip 0.2cm
\includegraphics[width=0.66\linewidth]{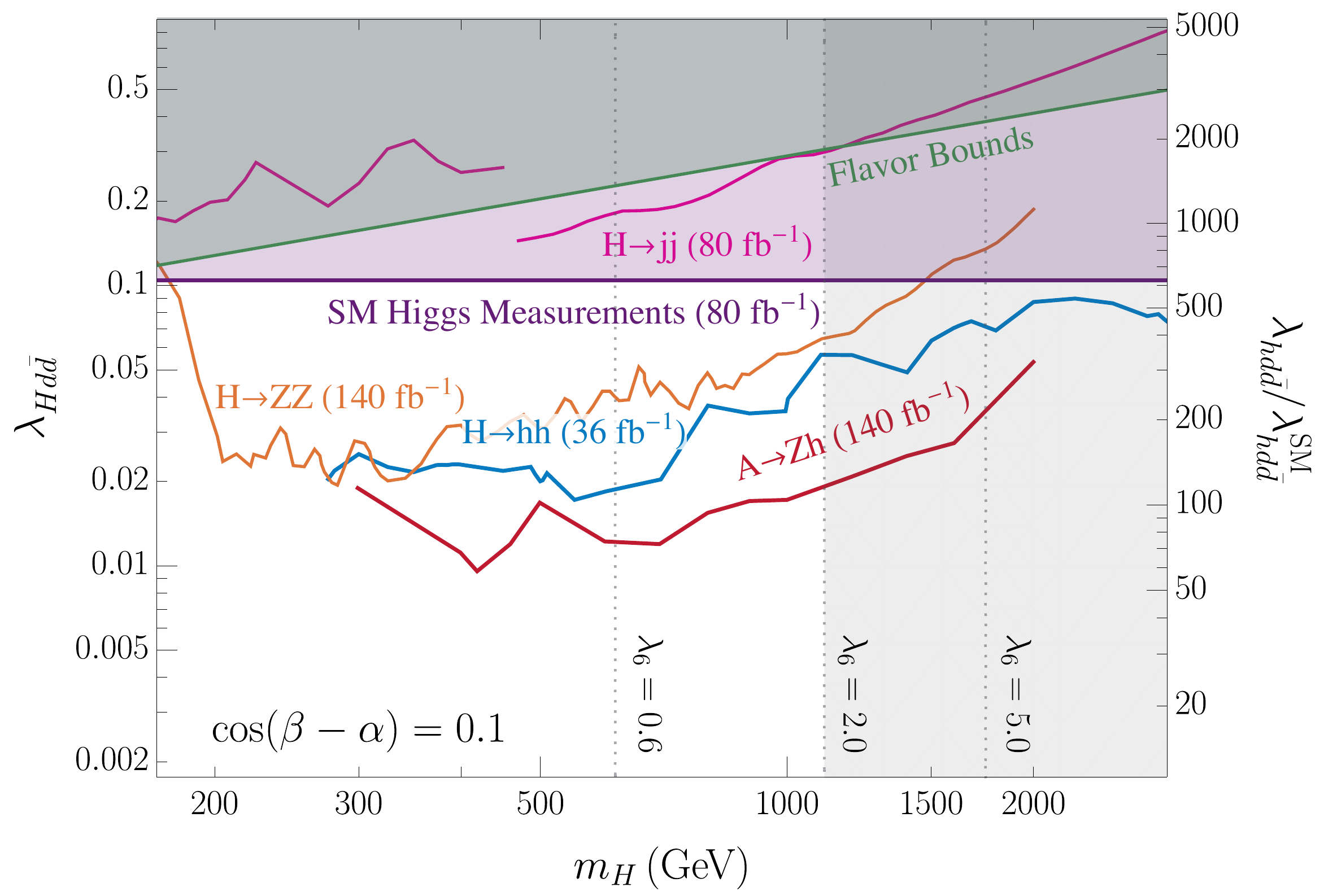}
\vskip -.25cm
\caption{
LHC bounds on a 2HDM with specific couplings to first-generation down quarks as a function of the heavy Higgs mass $m_H$ and its coupling to the down quark, $\lambda_{Hd\bar{d}}$ for fixed alignment parameter $\cos(\beta - \alpha) = 0.025$ (top) and $0.1$ (bottom).
On the right axis we show the corresponding enhancement of the $125\,\textrm{GeV}$ Higgs Yukawa coupling to the down quark obtained from mixing with the extra scalar Higgs.
The blue curve indicates the bounds from resonant di-Higgs production at the LHC~\cite{Sirunyan:2018ayu, Aad:2019uzh} while the red curves show the bounds on a resonance decaying to $Zh$~\cite{ATLAS:2020pgp}.
Bounds from $ZZ$~\cite{Aad:2020fpj, Sirunyan:2018qlb} and dijet~\cite{Aaboud:2019zxd,Aad:2019hjw,Sirunyan:2019vgj} resonant searches are shown as orange and pink curves, while $D-\bar{D}$ mixing and $125\,\textrm{GeV}$ Higgs signal strength bounds are shown as green and purple regions.
The vertical dotted lines indicate values of the quartic $\lambda_6$ required to obtain the alignment parameter for different masses.
We require $\lambda_6 < 2$ to avoid landau poles at the TeV scale.
}\label{fig:rate1}
\end{figure}

Due to the large multi-Higgs production rates that are obtained in 2HDMs with large couplings to light quarks, 
strong bounds on these theories can be set using current ATLAS \cite{Aad:2019uzh} and CMS \cite{Sirunyan:2018two} resonant di-Higgs searches.
No tri-Higgs searches are currently available, 
so we leave a discussion of the discovery potential of this topology for the next section, and here we concentrate on present bounds.

We derive and show limits on our theory in Fig.~\ref{fig:rate1}.
The limits are presented as a function of the extra Higgs boson mass $m_H$ and its coupling to the down quark $\lambda_{Hd\bar{d}}$, 
for alignment parameter $\cos(\beta-\alpha)=0.025$ (left panel) and $\cos(\beta-\alpha)=0.1$ (right panel). 
Given that the extra Higgs boson mixes with the SM Higgs, 
the $125\,\textrm{GeV}$ Higgs inherits an irreducible enhancement of its coupling to the down quark
(c.f. Table~\ref{t:yukawaup}). 
Thus, 
our limits can be alternatively interpreted as bounds on the Yukawas of the $125\,\textrm{GeV}$ Higgs to such quark, or on its enhancements with respect to the SM expectations, $\lambda_{hd\bar{d}}/\lambda_{hd\bar{d}}^{\textrm{SM}}$. 
We show these enhancements on the right vertical axes of the figures.
Before entering into the discussion of bounds, 
we see that in our theory the $125\,\textrm{GeV}$ Higgs Yukawa to the down quark can be strongly enhanced with respect to the SM expectations. 

The figures immediately demonstrate that the di-Higgs topology can set very strong, percent-level constraints on the couplings of the extra Higgs to the down quark.
To compare the reach of di-Higgs production with respect to other probes, 
in the same figures
we also present a series of additional limits from collider and flavor searches.
Limits from resonant production of $Zh$ mediated by the pseudoscalar Higgs $A$, taken from~\cite{ATLAS:2020pgp}, are shown in red.
Bounds from $ZZ$ and dijet resonance searches mediated by the extra Higgses are drawn in orange and pink.
In green, we show bounds from $D-\bar{D}$ mixing, taken from~\cite{Egana-Ugrinovic:2019dqu},
which in the model discussed here is mediated by the charged Higgs at one-loop.
Finally, large values of $\lambda_{Hd\bar{d}}$ are excluded since they are correlated with too-large enhancements of the $125\,\textrm{GeV}$ Higgs Yukawas to the down quark, $\lambda_{hd\bar{d}}$. 
This leads to a dilution of other SM Higgs decay modes, and thus to incompatibilities with SM Higgs rate measurements.
To estimate this limit, we simply require that the measured inclusive gluon-fusion cross section satisfies the bounds in \cite{Aad:2019mbh}, as was done in \cite{Egana-Ugrinovic:2019dqu}.
We present the excluded region in purple.

Comparing all bounds,
we immediately see that resonant di-Higgs production represents one of the leading probes of extra Higgses with large couplings to light quarks
for most of the plotted parameter space above the kinematic di-Higgs production threshold. 
While it appears that the $A \to Zh$ limits set stronger bounds than the $H \to hh$ searches, this is largely a result of $Zh$ bounds being taken from a search with almost four times as much luminosity as the most recent $hh$ bounds.
We expect that once higher luminosity $hh$ limits are released by ATLAS and CMS, the $hh$ topology will compete with or supersede the $Zh$ bounds.  

When interpreted as bounds on the Yukawa couplings of the $125\,\textrm{GeV}$ Higgs to the down quarks, 
the $hh$ and $Zh$ topologies set stringent bounds on the enhancements over the SM expectations.
While these limits vary for different $m_H$ and alignment parameter, 
we see that they can be as stringent as $\lambda_{hd\bar{d}}\lesssim 30 \times y_d^{\textrm{SM}}$ for $\cos(\beta-\alpha)=0.025$, and  $\lambda_{hd\bar{d}}\lesssim 60 \times y_d^{\textrm{SM}}$ for $\cos(\beta-\alpha)=0.1$.  
This provides an \textit{order of magnitude} improvement with respect to previous best bounds in the literature, 
which are at the level $\lambda_{hd\bar{d}}\lesssim 10^3 \times y_d^{\textrm{SM}}$ \cite{Kagan:2014ila,Soreq:2016rae}.
Note that over most of the parameter space shown in the figures, bounds on the Yukawa enhancements coming from the resonant production of $H$ are stronger than either the limits coming from the measurements of the $125\,\textrm{GeV}$ Higgs couplings or flavor.

The bounds from $hh$, $Zh$ or $ZZ$ production on the enhanced $125\,\textrm{GeV}$ Higgs Yukawas weaken at very high and low values of the extra Higgs mass.
However, in the large $m_H$ limit, large Yukawa enhancements cannot be obtained while retaining perturbativity.
More precisely, near the decoupling limit the Yukawa enhancements are of order $\lambda_{Hd\bar{d}}  \cos(\beta-\alpha) \sim \lambda_{Hd\bar{d}} \, \lambda_6 v^2/m_H^2$ (c.f. Eqs.~\eqref{eq:approxalign} and Table~\ref{t:yukawaup}). 
The only way to obtain large enhancements for large $m_H$, 
is to increase the quartic $\lambda_6$.
This can be seen in Figs.~\ref{fig:rate1} from the vertical vertical dotted lines,
which show contours of this quartic.
As discussed in the previous section, larger values of $\lambda_6$ lead to non-perturbative behavior in the form of Landau poles near the TeV scale.
Requiring $\lambda_6 \leq 2$ leads to a limit $\lambda_{hd\bar{d}}\lesssim 300 \times y_d^{\textrm{SM}}$ across all parameter space where di-Higgs production is kinematically accessible, 
which is still a significant improvement over previous constraints.
Moreover, even when $m_H$ falls below the di-Higgs threshold,  
resonant dijet and $ZZ$ searches together with $D-\bar{D}$ mixing limits still constrain our theory. 
Taking this into account, the limits on the enhanced Yukawas remain at the level of $\lambda_{hd\bar{d}}\lesssim 300 \times y_d^{\textrm{SM}}$ across all parameter space. 

At this point, it is worth commenting on the possibility of enhanced light-quark Higgs Yukawas in the context of the SM EFT, which has been the preferred framework in previous literature~\cite{Kagan:2014ila,Bishara:2016jga,Aaboud:2017xnb,Sirunyan:2018fmm,Aaboud:2018txb,Alasfar:2019pmn,Sirunyan:2020mds,Falkowski:2020znk}.
In this context, enhancements to the light-quark Yukawas can be obtained from the cubic Yukawa operator in Eq.~\eqref{eq:SMEFT}.
Within the SM EFT, the leading bounds on the Yukawa enhancements come from the $125\,\textrm{GeV}$ Higgs coupling fits, and are of the order of $\lambda_{hd\bar{d}}\lesssim 10^3 \times y_d^{\textrm{SM}}$ as pointed out above.
In this regard, we only point out that from the bounds in Fig.~\ref{fig:rate1}
it is clear that when the extra Higgses are beyond the reach of the LHC (so that the SM EFT can be safely used without being constrained by resonant searches), 
if one wants to obtain enhancements as large as $\lambda_{hd\bar{d}}\sim 10^3 \times y_d^{\textrm{SM}}$,
the theory has Landau poles near the TeV scale, 
as $\lambda_6$ is large. 
We thus conclude that, at least within the context of a 2HDM UV completion, the SM EFT is not an adequate tool to describe BSM physics in the first-generation $125\,\textrm{GeV}$ Higgs Yukawas at the LHC.

Enhancements to the Higgs Yukawas can also be obtained in other theories that do not lead to resonant di-Higgs production,
such as models with new TeV scale colored vector-like quarks~\cite{Alasfar:2019pmn,Bar-Shalom:2018rjs}.
In this case, the limits derived in this work on the enhanced Yukawas do not apply. 
However, such vector-like quarks suffer from bounds from direct searches for colored particles and also from flavor.
A careful analysis is needed in order to determine how much can the Higgs couplings be enhanced in these models, which is beyond the scope of this work.
Regarding the usefulness on an EFT description that UV completes into the vector-like quark model, 
it is again important to note that in order to obtain the large deviations in the Higgs light-quark Yukawas that can be currently measured, within perturbative theories, the vector-like quarks must be light.
This reinforces and generalizes our point, that an EFT description is not the appropriate framework to describe BSM physics in the $125\,\textrm{GeV}$ Higgs light-quark Yukawas.

Until now, we have concentrated only in theories where the extra Higgs doublet couples specifically to down quarks,
but we expect most of our conclusions to hold for theories where couplings to the up quark are preferred instead.
In particular, in the context of the 2HDM, we expect di-Higgs and $Zh$ bounds on the extra Higgses to provide the most stringent constraints on the $125\,\textrm{GeV}$ Higgs Yukawas to the up quark, when kinematically accessible.

\begin{figure}[ht!]
\includegraphics[width=0.65\linewidth]{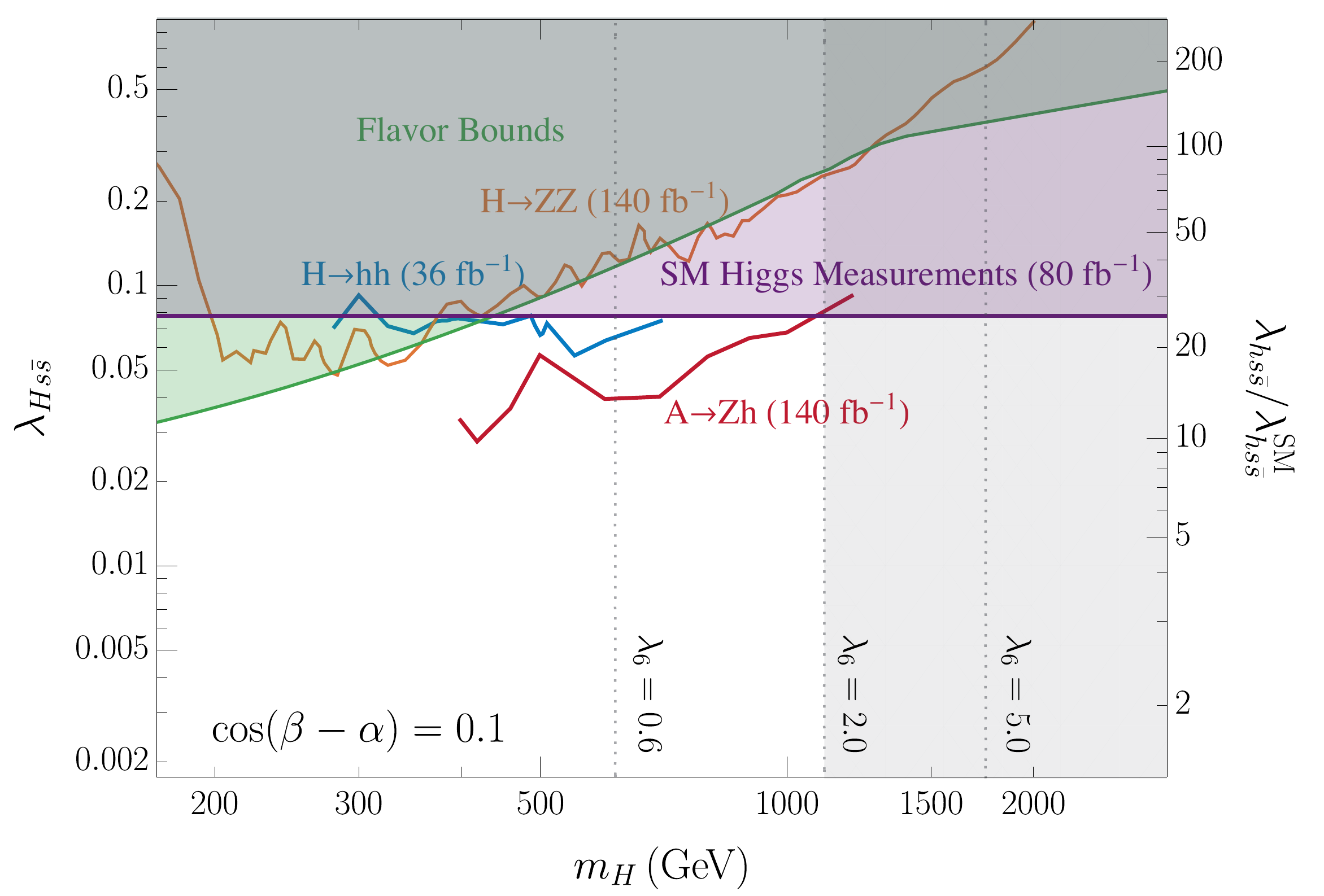}
\vskip -.25cm
\caption{
Contours of the LHC bounds on a 2HDM with specific couplings to the strange quarks as a function of the heavy Higgs mass $m_H$ and its coupling to down quarks, $\lambda_{Hd\bar{d}}$ for fixed alignment parameter $\cos(\beta - \alpha) = 0.1$.
On the right axis we show the corresponding enhancement of the $125\,\textrm{GeV}$ Higgs Yukawa coupling to strange quarks.
Bounds are presenting with the same color codings as in Fig.~\ref{fig:rate1}.
}\label{fig:ratestrange}
\end{figure}

Regarding specific couplings to strange quarks, 
in this case limits from di-Higgs and $Zh$ production are less spectacular due to the smallness of the strange-quark pdf, which leads to smaller $H$ resonant  production rates, but they are still a very stringent probe of enhanced strange Yukawas.
We show the corresponding bounds in Fig.~\ref{fig:ratestrange}.
From the figure, we clearly see that bounds from resonant $hh$ and $Zh$ production are already the most stringent bounds on enhanced strange Yukawas for $m_H \gtrsim 400\,$GeV,
and probe enhancements as small as $\lambda_{hs\bar{s}}\sim 10 \times y_s^{\textrm{SM}}$.
Moreover, these bounds will significantly improve at the HL-LHC, 
where we expect resonant di-Higgs production to represent the most stringent probe of the Yukawas of the $125\,\textrm{GeV}$ Higgs to the strange, in the context of the 2HDM.


\FloatBarrier

\section{Projections for the HL-LHC}
\label{sec:projections}

\begin{figure}[ht!]
\includegraphics[width=0.49\linewidth]{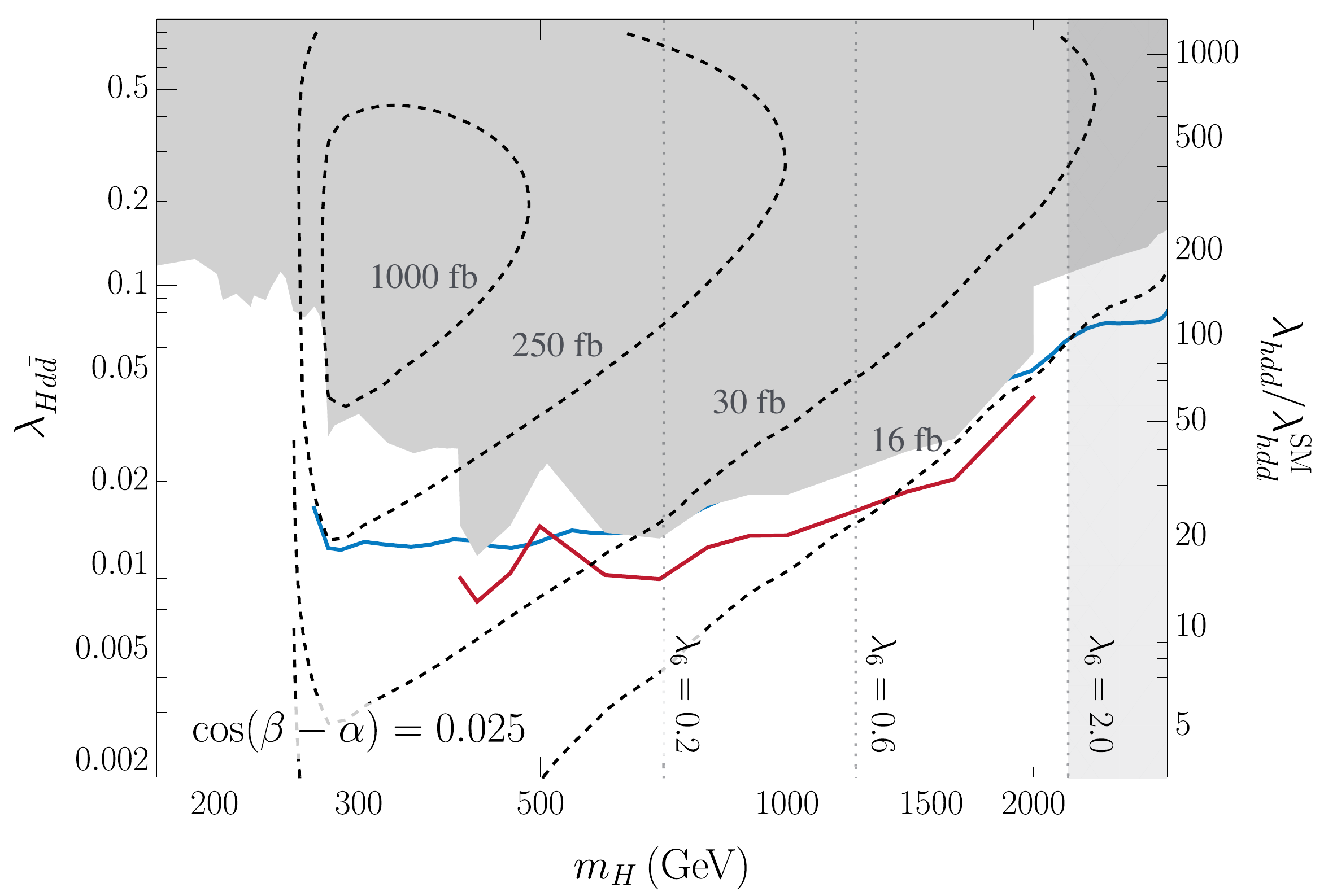}
\includegraphics[width=0.49\linewidth]{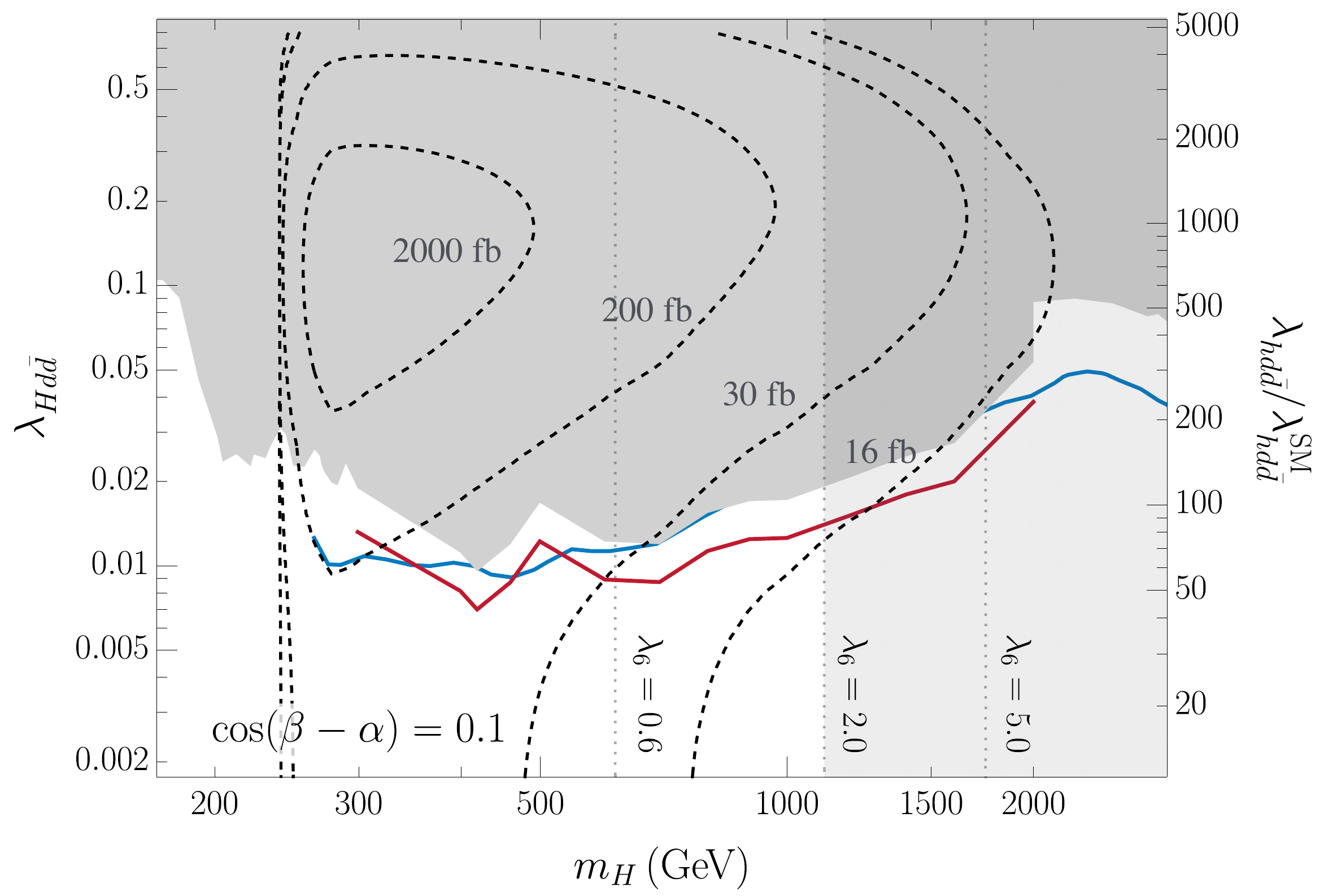}
\caption{
On the same axes as Fig.~\ref{fig:rate1}, 
we show in blue and red-solid lines the $5\sigma$ discovery reach for the HL-LHC for the 2HDM with specific couplings to first generation down quarks via the $H\to hh$ and $A \to Zh$ modes, respectively.
We also show contours of the di-Higgs production cross section, including the resonant contributions mediated by the extra Higgs $H$, as well non-resonant gluon-fusion contributions. 
The contours in this figure include the suppression of the SM Higgs branching ratios into the usual SM final states due to the enhanced couplings to the down quark. For comparison,  the NNLO gluon-fusion SM di-Higgs cross section at 13 TeV's is $\sigma_{hh} = 31.05\,\mathrm{fb}$.
The bounds from Fig.~\ref{fig:rate1} are shown collectively as a shaded gray region.}\label{fig:rate2}
\end{figure}

\begin{figure}[ht!]
\includegraphics[width=0.49\linewidth]{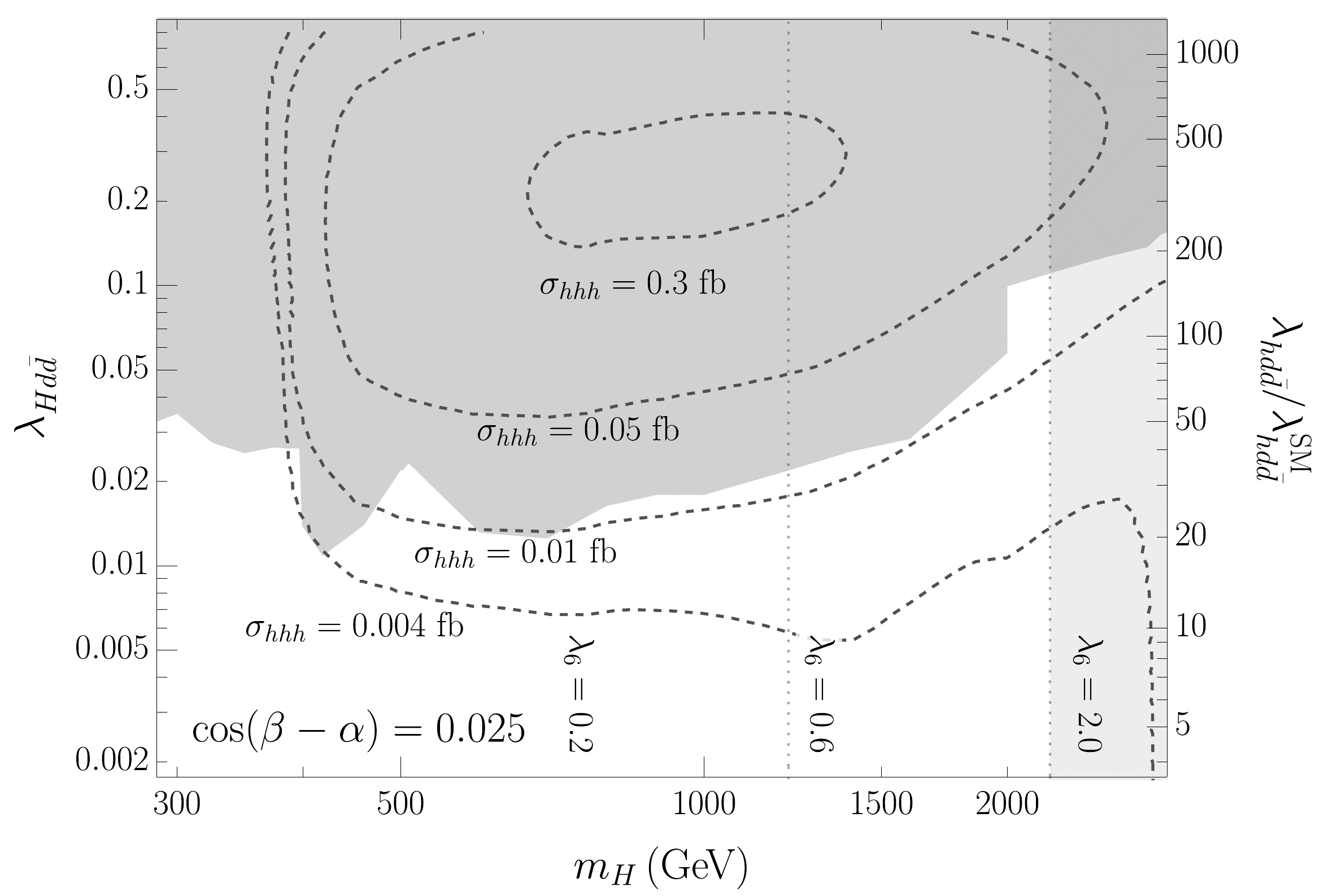}
\includegraphics[width=0.49\linewidth]{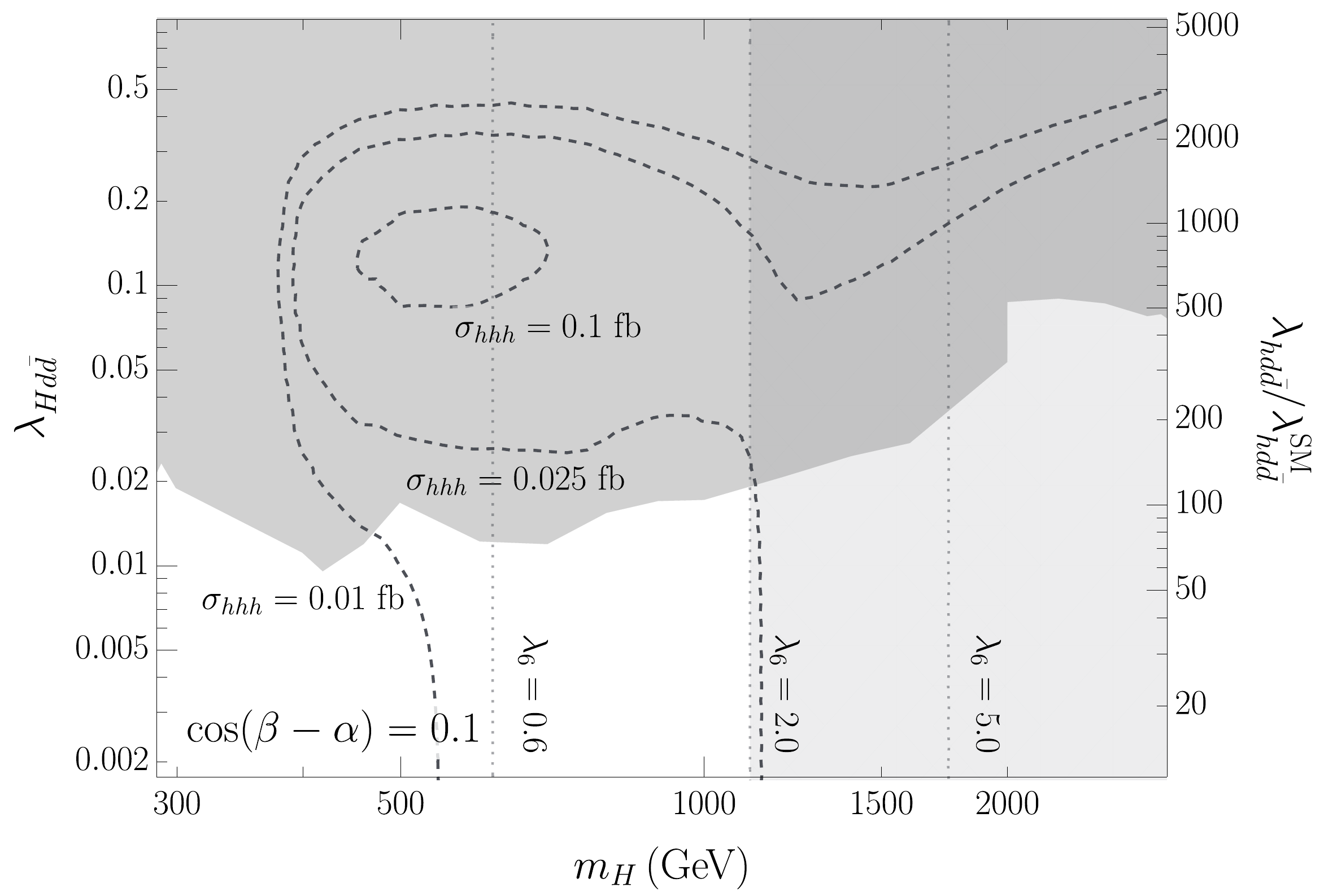}
\caption{
Contours of the $hhh$ production cross section, including the decay to six $b$'s and the kinematic acceptance from demanding all six $b$-jets to lie within $|\eta| < 2.5$ and have $p_{T,b} > 35\,\textrm{GeV}$, as a function of $m_H$ and $\lambda_{Hd\bar{d}}$ for $\cos(\beta - \alpha) = 0.025$ (left) and $\cos(\beta - \alpha) = 0.1$ (right).
The shaded regions are the same as in Fig.~\ref{fig:rate2}.
The kinematic acceptance is $\sim 25\%$ at $m_H = 750\,\textrm{GeV}$, and changes only modestly as a function of $m_H$.
}\label{fig:rate3}
\end{figure}

We now discuss the discovery potential of multi-Higgs production at the LHC in the context of our theories.
For concreteness, we again concentrate on the case of a 2HDM, allowing for specific couplings to the first-generation down-quarks.
We begin by studying di-Higgs production. 
In Fig.~\ref{fig:rate3}, we show in red and blue the HL-LHC ATLAS and CMS $5\sigma$ discovery reach for our 2HDM,
for $\cos(\beta-\alpha)=0.025$ (left panel), and $\cos(\beta-\alpha)=0.1$ (right panel). 
To obtain these projections, we simply assumed that at the HL-LHC the $hh$ backgrounds will increase as the square root of the luminosity with respect to current LHC background values.
From the figures we see that the HL-LHC has the potential to \textit{discover} enhanced $125\,\textrm{GeV}$ Higgs Yukawas to the down quark as small as $\lambda_{hd\bar{d}}\lesssim 15 \times y_d^{\textrm{SM}}$ for $\cos(\beta-\alpha)=0.025$, and $\lambda_{hd\bar{d}}\lesssim 50 \times y_d^{\textrm{SM}}$ for $\cos(\beta-\alpha)=0.1$. 
Such capabilities are unprecedented, and represent a dramatic improvement over any previously proposed test of enhanced Higgs Yukawas to any of the first generation SM fermions.

Moving on to the tri-Higgs topology, in Fig.~\ref{fig:rate3} we present contours of the tri-Higgs cross section times branching fraction into $b$-quarks, $\sigma(3h)\times (\Br(h\rightarrow bb))^3$ at the 13 TeV HL-LHC, 
for $\cos(\beta-\alpha)=0.025$ and $\cos(\beta-\alpha)=0.1$ on the left and right panels correspondingly.
In order to account for typical kinematic cuts that may be expected in a potential search for this topology, in the calculation of the cross section we have required all six $b$-jets to lie within $|\eta| < 2.5$ and have $p_{T,b} > 35\,\textrm{GeV}$. 
From the contours we see that after accounting for the branching fraction into $b$'s, kinematic acceptances, and consistent with current bounds on our 2HDM, 
more than $\sim 30$ resonant $3h$ events can be obtained at the HL-LHC for $\cos(\beta-\alpha)=0.025$, 
and twice as many for $\cos(\beta-\alpha)=0.1$.  
Given the rather spectacular final state, with 3 pairs of $b$'s each one with a reconstructed invariant mass equal to $m_h$, 
we expect backgrounds for this topology to be small. 
In particular, under the assumption of zero backgrounds, as few as three tri-Higgs events could be required at the HL-LHC to set $2\sigma$ bounds on our theory using this topology.
Then, from the figures we see that tri-Higgs production could be used to set stringent constraints on enhanced Higgs Yukawas to the first generation quarks, as strong as 
$\lambda_{hd\bar{d}}\lesssim 10 \times y_d^{\textrm{SM}}$ for $\cos(\beta-\alpha)=0.025$.

Even when the coupling of the extra Higgses to the first generation down quarks is negligibly small, as in the types I-IV 2HDMs or the singlet-extended SM, tri-Higgs production could be used to set stringent constraints on extra Higgses
since it can still arise due to $hH$ associated production with $H\rightarrow hh$,
as discussed in Section~\ref{sec:dihiggs}.
In fact, from the right panel in Fig.~\ref{fig:rate3}, we see that  even for $\lambda_{Hd\bar{d}}=0$, as many as 75 tri-Higgs events can be obtained within the 2HDM at the HL-LHC for $\cos(\beta-\alpha)=0.1$, 
consistent with current bounds and perturbativity requirements.
It is now clear that tri-Higgs production could represent one of the leading probes of light quark Yukawas and other generic extended Higgs sectors at the HL-LHC, providing strong motivation to develop a dedicated search for this topology.


\section{Phenomenological summary and conclusions}
\label{sec:conclusions}

In this work we studied the relevance of multiple-Higgs production for probing extended Higgs sectors at hadron colliders.
Regarding di-Higgs production and using a 2HDM as an example, we demonstrated that if extra Higgses have large couplings to light quarks, huge production rates of Higgs pairs are obtained that can be searched for at the LHC.
The rates are far above the ones obtained in popular extensions of the Higgs sector, 
such as the types I-IV 2HDMs and the singlet-extended SM~\cite{Bowen:2007ia,Dolan:2012ac,Chen:2014ask}.
In this context, we demonstrated that by using using resonant di-Higgs, $ZZ$ and $Zh$ searches at ATLAS and CMS,
strong bounds on the couplings of extra Higgses to the down quarks can be obtained.  
In addition, since the extra Higgs bosons mix with the SM Higgs, the latter inherits large couplings to light quarks. 
Thus, di-Higgs limits on our theories can be recast as limits on the enhancements of the $125\,\textrm{GeV}$ Higgs couplings to such quarks. 
In this regard, we found that constraints as stringent as $\lambda_{hd\bar{d}}\lesssim 30 \lambda_{hd\bar{d}}^{\textrm{SM}}$ can be {\em currently} set, well beyond any indirect or direct current constraints from single Higgs properties.
We expect significant improvements in these bounds with upcoming data from the next LHC run and the HL-LHC. 

Regarding tri-Higgs production, and as first pointed out by the authors of this work~\cite{SNOWMASS},
we found that large rates for this topology can be found in a variety of extended Higgs sectors.
In the context of a 2HDM that couples to first-generation quarks, we found that dozens of resonantly produced triple-Higgs events with decays into $6b$-jets can be obtained at the HL-LHC.
On the other hand, even when couplings to first-generation quarks are small so that resonant $qq$-initiated triple-Higgs production is not important, 
we demonstrated that large $3h$ rates can still be obtained due to 
associated production of an extra Higgs and a $125\,\textrm{GeV}$ Higgs, $hH$, followed by $H\rightarrow 2h$. 
In this way, we showed that tri-Higgs production could be important to explore generic extensions of the Higgs sector, such as the types I-IV 2HDMs and the singlet-extended SM. 

Based on this work, several additional avenues for progress exist. 
In relation to probes of Higgs flavor via multiple-Higgs production, 
while here we concentrated mostly on couplings of extra Higgses to the down-type quarks,
this analysis should be carried out for the up-type quarks too, 
including a detailed analysis of flavor bounds on UV completions in this scenario. 
We expect such flavor bounds to be more stringent than for the down-type quark case studied here due to strong constraints from $K-\bar{K}$ mixing, as discussed in~\cite{Egana-Ugrinovic:2019dqu}.

Second, while here we have discussed theories with large couplings to light quarks,
similar theories can be constructed leading to light couplings to light leptons.
It is rather trivial to extend the SFV flavor Ansatz to the lepton sector, 
by simply adding right-handed neutrinos to the SM,
and using the same mechanism as in \cite{Egana-Ugrinovic:2018znw}.\footnote{The mechanism in \cite{Egana-Ugrinovic:2018znw} introduces all flavor mixing to the theory via wave-function renormalization of right-handed quarks. For leptonic SFV, all flavor mixing to the lepton sector must be introduced then via wave-function renormalization of right-handed leptons.}
In these \textit{leptonic SFV theories}, one obtains a theory with Dirac neutrinos, 
where all BSM physics such as extra Higgs bosons is flavor-aligned, 
so they may have large couplings to light leptons without bumping into stringent bounds from leptonic FCNCs. 
In complete analogy to the studies done in this work, in leptonic SFV 2HDMs one would then naturally get large multi-Higgs production rates at \textit{future lepton colliders},
and huge modifications to the $125\,\textrm{GeV}$ Higgs couplings to light leptons.
The phenomenology of such theories remains unexplored, but presents several opportunities to look for BSM physics at colliders.
In addition, both quark-specific and lepton-specific SFV 2HDMs are especially relevant and motivated targets for future colliders, such as the FCC-hh and FCC-ee.
In particular, BSM physics with large couplings to light quarks provides an excellent motivation for developing light-flavor taggers~\cite{Duarte-Campderros:2018ouv,Nakai:2020kuu}.
Dedicated studies to understand the reach of such colliders to extended Higgs sectors with large couplings to light quarks or leptons are yet to be performed. 

Third, 
in regards to di-Higgs production, in this work we have mostly studied the resonant regime. 
New physics may also be discovered in the non-resonant di-Higgs topology, when production blind-spots are realized, \textit{i.e.}, when the extra Higgs states are light but they are not resonantly created due to smallness or cancellations in the couplings controlling the production. 
In this case, extra Higgses can avoid constrains by direct searches, but may still induce large deviations in the $125\,\textrm{GeV}$ Higgs trilinear coupling via mixing (differently from the situation when an EFT description applies and only small deviations can be obtained). 
These modifications can in turn lead to enhancements to the SM expectations for non-resonant di-Higgs production, which can be observed at the LHC.
Such blind-spots, which to our knowledge have not been explicitly pointed out in the literature, are one of the few realizations where measurements of the SM triple-Higgs coupling are relevant to set bounds on concrete and well-defined theories, so they deserve further study.

And finally, here we demonstrated that a comprehensive analysis of triple-Higgs production at the LHC, 
both as a probe of couplings of the Higgs to light quarks, 
and as a probe of conventional extended Higgs sectors such as the types I-IV 2HDMs and the singlet-extended SM, must be urgently carried out. 
While tri-Higgs production rates are obviously smaller than di-Higgs ones, they are still sizable, and backgrounds are expected to be small.
In addition,
triple-Higgs production has variety of rich kinematical features that can be exploited to further reduce backgrounds.

As we come closer to the HL-LHC being realized, we must remain mindful of the theoretical interpretations that its results may suggest, and of any gaps left in coverage. Here we have shown new targets and novel searches, which could bring spectacular discoveries in the coming years.

\section*{Acknowledgments}
DEU is supported by Perimeter Institute for Theoretical Physics. 
Research at Perimeter Institute is supported in part by the Government of
Canada through the Department of Innovation, Science
and Economic Development Canada and by the Province
of Ontario through the Ministry of Economic Development, Job Creation and Trade. 
DEU would like to thank the Simons Center for Geometry and Physics for the hospitality during part of the completion of this work.
The work of SH and PM was supported in part by the National Science Foundation grant PHY-1915093.
The work of SH was also supported in part by the DOE Grant DE-SC0013607 and by the Alfred P. Sloan Foundation Grant No. G-2019-12504
We would also like to thank Alain Blondel, Sally Dawson, Javier Duarte,  Philip Harris, Cristian Pe\~na and Si Xie for useful discussions.


\bibliographystyle{utphys}
\bibliography{qqdihiggs}

\end{document}